# TOWARDS EXPERIMENTAL TESTS OF QUANTUM EFFECTS IN CYTOSKELETAL PROTEINS

**Andreas Mershin\*[1,2], Hugo Sanabria[3], John H. Miller[3], Dharmakeerthna Nawarathna[3], Efthimios M.C. Skoulakis[1,4], Nikolaos E. Mavromatos[5], Alexadre A. Kolomenskii[1], Hans A. Schuessler[1], Richard F. Luduena[6], Dimitri V. Nanopoulos[1,7]**

[1] *Texas A&M University, Department of Physics, College Station, TX 77843-4242, USA*
[2] *Massachusetts Institute of Technology, Center for Biomedical Engineering, 77 Massachusetts Ave., Rm. NE47-376 Cambridge, MA 02139-4307, USA*
[3] *Dept. of Physics and Texas Center for Superconductivity, University of Houston Houston, TX 77204-5005, USA*
[4] *Institute of Molecular Biology and Genetics, Biomedical Sciences Research Centre, "Alexander Fleming" 34 Fleming St., Vari 16672, Greece*
[5] *Department of Physics, Theoretical Physics Group, University of London, King's College, Strand, London WC2R 2LS, U.K.*
[6] *Department of Biochemistry University of Texas Health Science Center at San Antonio San Antonio, TX 78229-3900, USA*
[7] *Academy of Athens, Natural Science Division, Athens, 10679, Greece*

## PREFACE

This volume is appropriately titled "The *Emerging* Physics of Consciousness" and much of it is focused on using some aspect of "quantum weirdness" to solve the problems associated with the phenomenon of consciousness. This is sometimes done in hopes that perhaps the two mysteries will somehow cancel each other through such phenomena as quantum coherence and entanglement or superposition of wavefunctions. We are not convinced that such a cancellation can take place. In fact, finding that quantum phenomena are involved in consciousness, what we will call the "Quantum Consciousness Idea" (QCI) (fathered largely by Penrose and Hameroff [1-3]), is likely to confound both mysteries and is of great interest.

In our contribution, we want to emphasize the "Emerging" part of this volume's title by pointing out that there is a glaring need for properly controlled and reproducible experimental work if any proposed quantum phenomena in biological matter, let alone consciousness are to be taken seriously.

There are three broad kinds of experiments that one can devise to test hypotheses involving the relevance of quantum effects to the phenomenon of consciousness. The three kinds address three different scale ranges associated roughly with tissue-to-cell (1cm-10μm), cell-to-protein (10μm-10nm) and protein-to-atom (10nm-1Å) sizes. Note that we are excluding experiments that aim to detect quantum effects at the "whole human" or even "society" level as these have consistently given either negative results or been plagued by irreproducibility and bad science (e.g. the various extra sensory perception and remote viewing experiments [4]).

The consciousness experiments belonging to the tissue-cell scale frequently utilize apparatus such as electro-encephalographs (EEG) or magnetic resonance imaging (MRI) to track responses of brains to stimuli. The best example of such is the excellent work undertaken by Christoff Koch's group at Caltech [5] sometimes in collaboration with the late Francis Crick [6], tracking the activity of living, conscious human brain neurons involved in visual recognition. These experiments are designed to elucidate the multi and single-cellular substrate of visual consciousness and awareness and are likely to lead to profound insights into the working human brain. Because of the large spatial and long temporal resolution of these methods, it is unclear whether they can reveal possible underlying quantum behavior (barring some unlikely inconsistency with classical physics such as for instance non-locality of neural firing).

---

\*To whom correspondence should be addressed: mershin@mit.edu

The second size scale that is explored for evidence of quantum behavior related to aspects of consciousness (memory in particular) is that between a cell and a protein. Inspired by QCI, seminal experimental work has been done by Nancy Woolf [7, 8] on dendritic expression of MAP-2 in rats and has been followed by significant experiments performed by members of our group on the effects of MAP-TAU overexpression on the learning and memory of transgenic *Drosophila* (summarized in Section 3). Such attempts are very important to the understanding of the intracellular processes that undoubtedly play a significant role in the emergence of consciousness but it is hard to see how experiments involving tracking the memory phenotypes and intracellular redistribution of proteins can show a *direct* quantum connection. It seems clear that experimentation at this size scale can at best provide evidence that is "not inconsistent with" and perhaps "suggestive of" the QCI [9]

The third scale regime is that of protein-to-atom sizes. It is well understood that at the low end of this scale, quantum effects play a significant role and it is slowly being recognized that even at the level of whole-protein function, quantum-mechanical (QM) effects may be of paramount importance to biological processes such as for instance enzymatic action [10] or photosynthesis [11]

In what follows, we give a brief overview of our theoretical QED model of microtubules and the extensive experimental work undertaken (belonging to the second and third size scales). We conclude by pointing towards directions of further investigation that can provide direct evidence of quantum effects in the function of biological matter and perhaps consciousness.

# SECTION 1:   INTRODUCTION

## 1.1    Overview

In our contribution to the "Emerging Physics of Consciousness" volume, we will report on our efforts to address, in several different ways, the hypotheses concerning the existence and function of subcellular, protein-based information-processing elements with a special emphasis on ways to test the QCI as it pertains to cytoskeletal proteins.

Our system of interest is the neural cytoskeleton and specifically the tubulin polymers known as microtubules (MTs). This work was originally motivated by quantum physics-based scenarios of brain function and consciousness that implicated these structures [1-3, 12-14]. As the vast majority of such proposals are purely speculative, with no experimental evidence mentioned or sought our first step was to show that at least some of the predictions of the more realistic and specific scenarios proposed, such as for instance the models put forth by D.V. Nanopoulos et al. [13, 15, 16], are in fact experimentally testable. In the first part of this contribution, we present the QED model of microtubules explored in [16] and showcase some of its predictions such as superefficient energy transport and the possibility of biological quantum teleportation.

Initially, this research concentrated on testing one prediction that was common to many of these exotic theories: that microtubules had a global role in neural function and were not merely the structural scaffolding of cells occasionally assisting in mobility and intracellular transport. To test this prediction, we decided to investigate what effects a minimal disruption of the microtubular network of neural cells has on memory. Memory was selected because certain models such as for instance the Guitar String Model (GSM) of the engram suggested an intracellular memory encoding and storage mechanism highly dependent on the precise stoichiometry of microtubules and microtubule associated proteins (MAPs) such as TAU [15] (see Section 2).

In Section 3, we show how this stoichiometry can be manipulated in genetically altered organisms whose memory can then be tested to reveal any effects. Thus we ascertained that *Drosophila* olfactory associative memory does indeed suffer when the equilibrium of MAPs and MTs is disturbed [9] suggesting that the neural cytoskeleton plays a *central* role in memory encoding and retrieval beyond those that were traditionally reported.

The next step was investigating the properties of tubulin and MTs in depth since once experimentally implicated in information manipulation, it was not unreasonable to ask whether these protein structures do indeed work as biological and perhaps quantum binary digits (*biobits* or *bioqubits*). The dielectric constant and electric dipole moment of tubulin were chosen for closer study since these featured prominently in information manipulation schemes by naturally occurring or fabricated MT networks. In order to check the validity of previous computer simulations first done by the group of Jack Tuszynski [17] and later replicated by our group (giving virtually identical results -[18]), an experimental determination of the dielectric constant of tubulin and MTs was needed. This was undertaken in the high frequency (optical) region (using refractometry and surface plasmon resonance) [18] and low frequency region (using dielectric spectroscopy –previously unpublished data and reported here in Section 4) and it was found that the rather high dipole moment suggested by simulation is reasonable.

As a result of these efforts, novel experimental techniques and theoretical models have been developed that point to several directions of further experimental research from molecular electronics and proteomics to actually at least one way of *directly* testing for quantum coherence and entanglement in biological matter (Section 5).

Finally, in Section 6 we summarize our results and attempt a unification of concepts.

## 1.2    Tubulin and Microtubules

Proteins are the ubiquitous living machines inside all cells performing everything from signal and energy transduction to movement, force generation and catalysis of reactions. It has also been suggested that information manipulation and storage is a possible role for proteins at least in the case of tubulin, microtubules and associated proteins.

### 1.2.1    Tubulin Biochemistry

Tubulin is a common polar protein found mainly in the cytoskeleton of eukaryotic cells and especially enriched in brain tissue. Many of its properties have been studied both experimentally and theoretically because of its importance in mitosis, its role as the building block of microtubules and its relevance to the pathology of several neurodegenerative diseases including cancer and Alzheimer's.

Although the structure of tubulin has been solved to better than 3.7Å, by electron crystallography [19, 20] yielding data suggesting that the tubulin heterodimer has dimensions 46x80x65 Å (Fig. 1 (b)), the dipole moment has so far only been calculated via computer simulations [17, 18].

Microtubules (MTs) are hollow (25nm-outer diameter, 14nm inner diameter) tubes (Fig. 1 (a)); they constitute a major portion of the cytoskeleton. Apart from giving shape and support to the cell, they also play a major role in cellular transport and have been hypothesized to be central in cellular information processing. The structure of MTs has been the subject of comprehensive study using electron and optical microscopy. The structure of the microtubule is 13-protofilament B-lattice. The helical pitch is 3 monomers per turn and there is a "seam" where adjacent protofilaments join. The microtubule is thus an asymmetric structure [21, 22]. Under normal physiological conditions, tubulin is found as a heterodimer, consisting of two nearly identical monomers called α- and β- tubulin, each of molecular weight of about 50 kDalton [20]. GDP-GTP exchange (hydrolysis) releases approximately 0.42eV per molecule and can be modeled by a conformational change resulting in a 27° angle [23] between the original line connecting the centers of the α and β monomers and the new center-to-center line as shown in Fig. 1 (d). Note that for free tubulin, the energy needed for this conformational change is roughly 200 times lower than a conventional silicon-based binary switch and about 30 times more than thermal noise at room temperature. Therefore, at least energy-wise, these two conformations can act as the basis for a naturally occurring or fabricated biobit.

There exists a large number of studies dealing with microtubule (MT) dynamics, and various scenarios have been proposed for explaining the dynamic behavior of MTs. At this point however, there is controversy even as to the correct mechanism of polymerization with the "GTP cap" theory [24]facing alternatives such as those described in [25].

Unless otherwise specified, we will refer to the αβ-dimer simply as tubulin. Tubulin binds to two molecules of guanosine 5'-triphosphate (GTP). One GTP binds *irreversibly* to the α subunit; this is referred to as the "non-exchangeable" GTP; We will not concern ourselves further with this GTP. The β subunit binds *reversibly* to another GTP molecule; this is referred to as the "exchangeable" GTP. If β is bound to GTP then the tubulin dimer exists in an energy-rich form that favors polymerization. Alternatively, β can bind guanosine 5' diphosphate (GDP-tubulin) in which case the dimer would be in an energy-poor form (GDP-tubulin) that favors dissociation [26, 27]. Above 0° C, free tubulin can self-assemble into MTs in vitro provided the buffer contains sufficient GTP and the concentration is above critical (about 1mg/mL) so that sufficient nucleation sites exist. In tubulin from mammalian brains (the most commonly used source, and the one we have used in the experiments of Section 4) MTs do not assemble at 4 °C while they start assembling at around 17 C. The lower the temperature, the more they disassemble, until by 4 °C there are no microtubules left. Antarctic fish on the other hand have MTs that can assemble even at -1°C). The size of the minimum nucleus required to start polymerization is not exactly understood.

Certain interesting phenomena arise in microtubules such as length oscillations (known as dynamic instability), or treadmilling; these have been studied extensively [28, 29] but are not directly relevant to our analysis at this stage as such phenomena can be avoided in vitro by choosing appropriate environments.

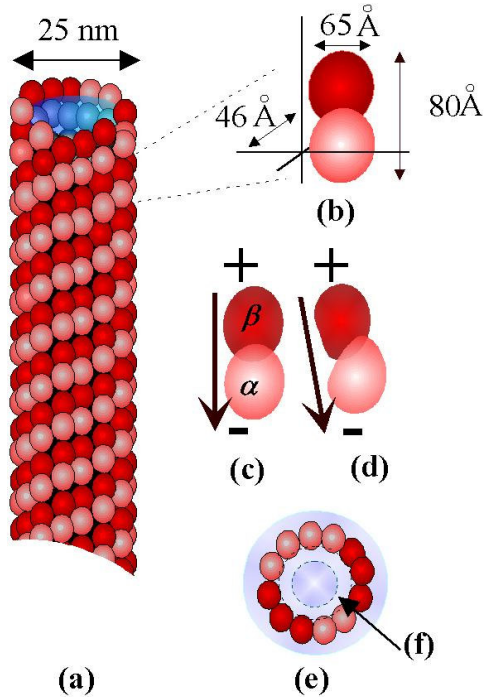

**Fig. 1 Microtubules and Tubulin.** (Taken from [16]). MTs are hollow (averaging 25nm-outer diameter, 14nm inner diameter) tubes forming the main component of the cytoskeleton **(a)** Typical microtubule made of 13 tubulin protofilaments. **(b)** dimensions of the aβ-heterodimer. **(c)** GTP-tubulin. **(d)** GDP-tubulin **(e)** a cross-section of the MT showing water environment **(f)** thin isolated region that has been theoretically suggested to possibly be equivalent to a quantum optical (QED) cavity [30].

### 1.2.2 Tubulin Biophysics

A measurement of the tubulin electric dipole moment will be useful in simulations of MTs that aim at understanding the polymerization mechanism, as it can be incorporated into the various models as an experimentally determined parameter. For similar reasons, computer simulation of MT networks as cellular automata will also benefit from such a measurement. Furthermore, drug interactions with tubulin are currently under investigation and it has been theorized that electric dipole moment 'flips' are responsible for London forces during interaction of tubulin with other molecules possessing dipole moments such as general anesthetic molecules [3].

A simplistic estimate of the tubulin dipole moment $p$ based on a mobile charge of 18 electrons multiplied by a separation of 4nm gives a magnitude of $p = 4 \times 10^{-27}$ C m (or 1200 Debye) while using a more sophisticated molecular simulation, $p$ has been quoted at around 1700 Debye [17, 18] . At physiological pH (=7.2) MTs are negatively charged [23, 31, 32] due to the presence of a 15-residue carboxyl terminus 'tail' and there have been suggestions that this C-terminus is important in polymerization, protein interactions and perhaps charge conduction [32].

This had not been included in the original electron crystallography data of Nogales and Downing [20] so all values concerning the dipole moment are quoted with the understanding that $p$ has been calculated ignoring the effect of the C-terminus. It is expected that post-translational modifications such as polyglutamylation, polyglycylation, tyrosination/detyrosination, deglutamylation, and phosphorylation (reviewed in [33]) will affect the electrostatics and electrodynamics of tubulin and MTs. For instance, the addition of 6 glutamates to the C-terminus of both α and β makes this region extraordinarily negatively charged. It is also known that at pH 5.6 MTs become neutral. Finally, there have been some preliminary experiments aimed at measuring the electric field around MTs [34-36] indicating that MTs could be ferroelectric.

Apart from the above observations, there exists little experimental evidence concerning tubulin's and MTs' electrical properties. On the other hand, there have been a large number of publications of theoretical work describing various electrical and optical properties that tubulin and MTs are expected to have based on their structure and function. Ferroelectricity (spontaneous abrupt orientation of dipoles for an above-threshold externally applied electric field) has been thoroughly explored and so far supported by the analysis in [12]. The MTs' paracrystalline geometrical structure has been implicated in error-correcting codes [37]. Energy loss-free transport along MTs has also been theoretically shown in [14] and the presence of "kink like excitations" or solitons has also been suggested as an energy-transfer mechanism in MTs [16].

Many models exist regarding the exact nature of these excitations and waves but they all depend on the dipole moment of tubulin and its ability to flip while in the polymerized state so we will collectively call them "flip waves". Depending on the model and the parameters assumed, the speed of such waves has been estimated to be $10^{2\pm1}$m/s. Note that although biochemically it seems MTs are made predominantly of GDP-tubulin, this does not mean that the *dipole* flips are 'frozen-out' since intramolecular electron motion can still occur in the hydrophobic pockets of the tubulin heterodimer. Detection of such waves would be a major step towards understanding the function of MTs and also towards using them as nanowires for bioelectronic circuits. We suggest several ways to test for the existence of flip waves in Section 5.

### 1.3     Motivation
### 1.3.1   Tubulin and Classical Molecular Electronics

Recent efforts have concentrated on identifying various chemical substances with appropriate characteristics to act as binary switches and logic gates needed for computation. While at present the size of the smallest conventional silicon-based devices is around 180nm, molecular devices promise a one or two order of magnitude reduction in this minimum. For instance, rotaxanes have been considered as switches and/or fuses [38] and carbon nanotubes as active channels in field effect transistors [39]. Many of these are not suitable for placement on traditional chips [40] or for forming ordered networks while virtually all of these attempt to hybridize some kind of electrical wires to chemical substrates in order to obtain current flows. This requirement adds a level of complexity to the task because of the need for appropriate nanomanufactured wires and connections. The work presented here is suggestive of a different approach. The role of the binary states can be played not by the presence or bulk movement of charge carriers but by naturally occurring conformational states of protein (tubulin) molecules. The external interaction with these states may be performed by coupling laser light to specific spots of the MT network. Signal propagation can be achieved by traveling electric dipole moment flip waves along protofilaments and MTs while modulation can be achieved by MAP binding that creates "nodes" in the MT network. In this proposed scheme for information manipulation, there is no bulk transfer of charge or mass involved. Permanent information encoding can be achieved by creating geometrical arrangements via placing the MTs on chips spotted with appropriate chemicals (e.g. taxol, zinc, colchicine) to force specific features such as centers, sheets, spirals or elongated structures to emerge from the MT self-assembly.

The end products of tubulin polymerization can be controlled by temperature and application of chemicals and MAPs to yield closely or widely spaced MTs, centers, sheets, rings and other structures [41, 42] thus facilitating fabrication of nanowires, nodes and networks.

Charge conduction by peptides/natural switches has been under investigation using femtosecond laser pulses and the results show that electron transfer exists in peptides even at low energies [43]. Femtosecond processes are taking place as charges jump across peptide backbones in certain favorable conformations that provide efficient conduction, or charges are stopped by those same peptides assuming a non-conducting conformation [44]. These general properties of peptides make proteins even more attractive as molecular switch candidates. Finally, these assemblies of protein are ideal for implementing fault-tolerant operation, as massive redundancy is possible due to the relative ease of obtaining large amounts of protein polymerized to specifications, e.g. on etched chips.

### 1.3.2    Quantum Computing and Quantum Consciousness Idea (QCI)

At the core of many "quantum brain" hypotheses lies a QCI that assumes that the tubulin electric dipole moment is capable of flips *while in the polymerized state* and starting from that assumption, predictions have been developed such as long-lived superposed and entangled states among tubulin dimers [16] and long-range non-neurotransmitter based communication among neurons [3]. Interaction between water molecule dipoles and the tubulin dipole plays a central role in the models predicting ferroelectricity along MT protofilaments, emission of coherent photons, intracellular quantum teleportation of dipole quanta states [16] and other controversial yet fascinating features. As explained in Section 5, a minimally modified design of the surface plasmon resonance (SPR) apparatus built and used for the work presented here, coupled to a entangled photon source and a correlation device might be capable of detecting the often conjectured mesoscopic (of order micrometers) bulk coherence and partial quantum entanglement of dipole moment states, existence of which will cast biomolecules as appropriate candidates for the implementation of bioqubits.

**SECTION 2:    QED MODEL OF TUBULIN AND ITS IMPLICATIONS**

## 2.1    Introduction
### 2.1.1    Background

Observable quantum effects in biological matter such as proteins are routinely expected to be strongly suppressed, mainly due to the macroscopic nature of most biological entities as well as the fact that such systems live at near room temperature. These conditions normally result in a very fast collapse of the pertinent wave functions to one of the allowed classical states. However, we suggest that under certain circumstances it is in principle possible to obtain the necessary isolation against thermal losses and other environmental interactions, so that meso- and macroscopic quantum-mechanical coherence, and conceivably entanglement extending over scales that are considerably larger than the atomic scale, may be achieved and maintained for times comparable to the characteristic times for biomolecular and cellular processes.

In particular, it has been shown [14] how MTs [45] can be treated as quantum-mechanically isolated (QED) cavities, exhibiting properties analogous to those of electromagnetic cavities routinely used in quantum optics [46-48],[49]. Recently, this speculative model has been supported by some indirect experimental evidence. It has been experimentally shown [50], that it is possible to maintain partial entanglement of the bulk spin of a macroscopic quantity of Cesium (Cs) atoms (N=$10^{12}$), at room temperature, for a relatively long time (0.5ms). Note that in this experiment, the large quantity of atoms was of paramount importance in creating and maintaining the entanglement, and even though the gas samples were in constant contact with the environmental heat bath, Julsgaard et al. managed to detect the existence of entanglement for a much longer time than one would intuitively expect. We stress that it was *exactly the large amount of spins* that was responsible for the long decoherence times. Fewer spins would have decohered faster. Here, we use the main features of the Nanopoulos et al. QED model of the quantum mechanical properties of MTs (described in detail in [[16, 30]), and we exhibit its relevance to the Julsgaard et al. experiment. A direct consequence of this model for MTs as QED cavities is that virtually every experimentally known QED-cavity-based observation may have an analogue in living MTs and we show this analytically with the specific case of intra- and inter-cellular dissipation-less energy transfer and quantum teleportation of coherent (biologically relevant) quantum states.

### 2.1.2    Intracellular Energy Transfer

Energy transfer across cells, without dissipation, was first speculated to occur in biological matter by Fröhlich [51]. The phenomenon conjectured by Fröhlich was based on a one-dimensional superconductivity model: a one dimensional electron system with holes, where the formation of solitonic structures due to electron-hole pairing results in the transfer of electric current without dissipation. Fröhlich suggested that, if appropriate solitonic configurations are formed inside cells, energy in biological matter could also be transferred without any dissipation (superefficiently). This idea has lead theorists to construct various models for cellular energy transfer, based on the formation of kink classical solutions [52].

In these early works, no specific microscopic models had been considered. In 1932 Sataric et al. suggested a classical physics model for microtubule dynamics [53], in which solitons transfer energy across MTs without dissipation. In the past, quantum aspects of this one-dimensional model have been analyzed, and a framework for the consistent quantization of the soliton solutions was developed [30]. That work suggested that such semiclassical solutions may emerge as a result of 'decoherence' due to environmental interactions, echoing ideas in [54] .

The basic assumption used in creating the model of [53] was that the building blocks of MTs, the tubulin molecule dimers, can be treated as elements of Ising spin chains (one-space-dimensional structures). The interaction of each tubulin chain (protofilament) with the neighboring chains and the surrounding water environment was accounted by suitable potential terms in the one-dimensional Hamiltonian. The model describing the dynamics of such one-dimensional sub-structures was the ferroelectric distortive spin chain model of [53].

Ferroelectricity is an essential ingredient of the quantum-mechanical mechanism of energy transfer that we propose. It has been speculated [12] that the ferroelectric nature of MTs will be that of hydrated ferroelectrics, i.e. the ordering of the electric dipole moment of the tubulin molecules will be due to the interaction of the tubulin dimers' electric dipoles with the water molecules in the interior and possibly exterior of the microtubular cavities. Ferroelectricity induces a dynamical dielectric 'constant' $\varepsilon(\omega)$ which is dependent on the frequency $\omega$ of the excitations in the medium. Below a certain frequency, such materials are characterized by almost vanishing dynamical dielectric 'constants', which in turn implies that electrostatic interactions inversely proportional to $\varepsilon$ will be enhanced, and thus become dominant against thermal losses. In the case of microtubules, the pertinent interactions are of the electric dipole type, scaling with the distance r as $1/(\varepsilon r^3)$. For ordinary water media, the relative dielectric constant $\kappa$ $(=\varepsilon/\varepsilon_o)$ is of order 80. In the ferroelectric regime however, $\varepsilon$ is diminished significantly. As a result, the electric dipole- electric dipole interactions may overcome the thermal losses that are proportional to $k_B T$ (where $k_B$ is Boltzmann's constant $= 1.38 \times 10^{-23}$ J/K and T is the temperature in degrees K) at room temperature inside the interior cylindrical region of MT bounded by the dimer walls of thickness of order of a few Å [30] see Fig. 1.

### 2.1.3   Cavities

Once isolation form thermal noise is provided, one can treat the thin interior MT regions as electromagnetic cavities in a way similar to that of QED cavities. Note that the role of MT as waveguides has been proposed by S. Hameroff already some time ago [55]. In this scenario on the other hand, we are interested in isolated regions inside the MT that play the role of QED cavities not waveguides.

QED cavities are well known for their capacity to sustain in their interior coherent modes of electromagnetic radiation. Similarly, one expects that such coherent cavity modes will occur in the thin interior regions of MTs bounded by the protein dimer walls. Indeed, as was discussed in [30], these modes are provided by the interaction of the electric dipole moments of the ordered-water molecules in the interior of MT with the quantized electromagnetic radiation [56, 57]. Such coherent modes are termed dipole quanta. It is the interaction of such cavity modes with the electric dipole excitations of the dimers that leads to the formation of coherent (dipole) states on the tubulin dimer walls of MTs. A review of how this can happen, and what purely quantum effects can emerge from the QED nature of MTs, will be the main topic of this Section.

A concise exposé of the mechanism described in detail in  [12, 14, 16, 30] is presented here that justifies the application of QM to the treatment of certain aspects of MT dynamics. An analogy of this mechanism to the experimental setup used by Julsgaard et al. [50] is drawn. A straight-forward calculation of how quantum teleportation of states can occur in MTs, in direct analogy to the experimental quantum teleportation in optical cavities that has been observed recently [58, 59] is performed. A parallel between certain geometrical features of MTs such as their ordered structure which obeys a potentially information-encoding code is illustrated and suggestions to exploit for this for (quantum) error-correction and dense coding are put forth.

## 2.2     Quantum Coherence in Biological Matter ?
### 2.2.1    Tubulin, Microtubules and Coherent States

Tubulin and MTs have been described in detail in Section 1. The interior of the MT, seems to contain ordered water molecules [30], which implies the existence of an electric dipole moment and an electric field. We stress here that the intracellular ordered water which is full of proteins and other molecules is different from ordinary water in various respects e.g. as is implied in [60]. It has been put forward that each dimer has two hydrophobic pockets, containing 2×18 unpaired electrons [45] that have at least two possible configurations associated with the GTP and GDP states of tubulin, which we will call ↑ and ↓ electron (or equivalently electric dipole moment) conformations respectively.

It is evident that an experimentally determined electric dipole moment for the tubulin molecule and its dynamics are important areas of study for this field. If we account for the effect of the water environment that screens the electric charge of the dimers by the relative dielectric constant of the water, $\kappa \sim 80$, we arrive at a value

$$p_{dimer} = 3 \times 10^{-28} C \times m \tag{1}$$

Note that under physiological conditions, the unpaired electric charges in the dimer may lead to even further suppression of $p_{dimer}$.

Note that although the caps of the MT contain both GTP and GDP tubulin, it is well known experimentally [23] that the tubulin comprising the trunk of the MT is GDP-tubulin incapable of acquiring a phosphate and becoming GTP tubulin. However, this does not preclude electric-dipole moment flip wave propagation down the MT, as a flip at the cap can be propagated without phosphorylation but rather via the mechanism suggested below. In view of this, the value of the yet undetermined electric dipole moment direction flip angle $\theta_{flip}$ is much smaller than the $27^{o}42'$ value for free tubulin and it is entirely unphysical to suggest a complete reversal of the dipole moment direction as some authors have recently done. The ↑ and ↓ states still exist, but are hard to observe experimentally as they are not associated with a large-scale geometrical mass shift.

Note that virtually all of the MT-based QCIs today fail to take this into account and instead wrongly suggest that $\theta_{flip}$ is of the order of $27^{o}$ and that such large distortions occur in the trunk of the polymerized MT.

In standard models for the simulation of MT dynamics [53], the physical degree of freedom which is relevant for the description of energy transfer is the projection of the electric dipole moment on the longitudinal symmetry axis (x-axis) of the MT cylinder. The $\theta_{flip}$ distortion of the ↓-conformation leads to a displacement $u_n$ along the x-axis. This way, the effective system is one-dimensional (spatial), and one has the possibility of being quantum integrable [30].

It has been suggested for quite some time that information processing via interactions among the MT protofilament chains can be sustained on such a system, if the system is considered as a series of interacting Ising chains on a triangular lattice. For such schemes to work, one must first show that the electromagnetic interactions among the tubulin dimers are strong enough to overcome thermal noise. It is due to this problem that such models for intra-neuronal information processing have been criticized as unphysical [61]. Classically, the various dimers can only be in the ↑ and ↓ conformations. Each dimer is influenced by the neighboring dimers resulting in the possibility of a transition. This is the basis for classical information processing, which constitutes the picture of a (classical) cellular automaton.

If we assume (and there is theoretical basis for such an assumption [30]) that each dimer can find itself in a QM *superposition* of ↑ and ↓ states, a quantum nature results. Tubulin can then be viewed as a typical two-state quantum mechanical system, where the dimers couple to conformational changes with $10^{-9}$-$10^{-11}$ sec transitions, corresponding to an angular frequency $\omega \sim 10^{10}$-$10^{12}$ Hz. In this approximation, the upper bound of this frequency range is assumed to represent (in order of magnitude) the characteristic frequency of the dimers, viewed as a two-state quantum-mechanical system:

$$\omega_o \sim (10^{12}) \text{ Hz} \tag{2}$$

Let $u_n$ be the displacement field of the $n^{th}$ dimer in a MT chain. The continuous approximation proves sufficient for the study of phenomena associated with energy transfer in biological cells, and this implies that one can make the replacement

$$u_n \rightarrow u(x,t) \tag{3}$$

with x a spatial coordinate along the longitudinal symmetry axis of the MT. There is a time variable t due to fluctuations of the displacements u(x) as a result of the dipole oscillations in the dimers.

The effects of the neighboring dimers (including neighboring chains) can be phenomenologically accounted for by an effective potential V(u). In the model of ref. [53] a double-well potential was used, leading to a classical kink solution for the u(x,t) field. More complicated interactions are allowed in the picture of ref. [30] where we have considered more generic polynomial potentials.

The effects of the surrounding water molecules can be accounted for by a viscous force term that damps out the dimer oscillations,

$$F = -\gamma \partial_t u \tag{4}$$

with $\gamma$ determined phenomenologically at this stage. This friction should be viewed as an environmental effect, which however does not lead to energy dissipation, as a result of the non-trivial solitonic structure of the ground state and the non-zero constant force due to the electric field. This is a well-known result, directly relevant to energy transfer in biological systems [52].

The effective equation of motion for the relevant field degree of freedom u(x,t) reads:

$$u''(\xi) + \rho u'(\xi) = P(u) \tag{5}$$

where $\xi = x - vt$, v is the velocity of the soliton, $\rho$ is proportional to $\gamma$ [53], and P(u) is a polynomial in u, of a certain degree, stemming from the variations of the potential V(u) describing interactions among the MT chains [30]. In the mathematical literature [62] there has been a classification of solutions of equations of this form. For certain forms of the potential [30] the solutions include kink solitons that may be responsible for dissipation-free energy transfer in biological cells [52]:

$$u(x,t) \sim c_1(\tanh[c_2 (x-v \ t)] + c_3) \tag{6}$$

where $c_1, c_2, c_3$ are constants depending on the parameters of the dimer lattice model. For the form of the potential assumed in the model of [53], there are solitons of the form

$$u(x,t) = c_1' + \frac{c_2' - c_1'}{1 + e^{c_3'(c_2' - c_1')(x - vt)}}, \text{ where again } c_i' \text{ where i = 1,2,3 are appropriate constants.}$$

A semiclassical quantization of such solitonic states has been considered in [30]. The result of such a quantization yields a modified soliton equation for the (quantum corrected) field $u_q(x,t)$ [63]

$$\partial_t^2 u_q(x \cdot t) - \partial_x^2 u_q(x,t) + M^{(1)}[u_q(x,t)] = 0 \tag{7}$$

with the notation $M^{(n)} = e^{\frac{1}{2}(G(x,x,t) - G_0(x,x))\frac{\partial^2}{\partial z^2}} U^{(n)}(z)\Big|_{z=u_q(x,t)}$ and $U^n \equiv d^n U/dz^n$ where the quantity

$U$ denotes the potential of the original soliton Hamiltonian, and G (x,y,t) is a bilocal field that describes quantum corrections due to the modified boson field around the soliton. The quantities $M^{(n)}$ carry information about the quantum corrections. For the kink soliton (eq. 6) the quantum corrections (7) have been calculated explicitly in ref. [63], thereby providing us with a concrete example of a large-scale quantum coherent state.

A typical propagation velocity of the kink solitons (e.g. in the model of ref. [53]) is v ~ 2 m/sec, although, models with v ~ 20 m/sec have also been considered [64]. This implies that, for moderately long microtubules of length L ~$10^{-6}$m, such kinks transport energy without dissipation in

$$t \sim 5 \times 10^{-7} \text{sec} \qquad (8)$$

Energy will be transferred super-efficiently via this mechanism only if the decoherence time is of the order of, or longer than, this time. We shall see in fact that indeed such time scales are order comparable to, or smaller than, the decoherence time scale of the coherent (solitonic) states $u_0(x,t)$. This then implies that fundamental quantum mechanical phenomena may be responsible for frictionless, dissipationless super-efficient energy (and signal) transfer and/or transduction across microtubular networks in the cell.

### 2.2.2 Microtubules as Cavities

In [30], a microscopic analysis of the physics underlying the interaction of the water molecules with the dimers of the MT was presented. This interaction is responsible for providing the friction term (4) in the effective (continuum) description. We briefly review this scenario here.

As a result of the ordered structure of the water environment in the interior of MTs, there appear collective coherent modes, the so-called dipole quanta [56]. These arise from the interaction of the electric dipole moment of the water molecule with the quantized radiation of the electromagnetic field [57], which may be self-generated in the case of MT arrangements [30, 64]. Such coherent modes play the role of 'cavity modes' in the quantum optics terminology. These in turn interact with the dimer structures, mainly through the unpaired electrons of the dimers, leading to the formation of a quantum coherent solitonic state that may extend even over the entire MT network. As mentioned above, such states may be identified [30] with semi-classical solutions of the friction equations (5). These coherent, almost classical, states should be viewed as the result of decoherence of the dimer system due to its interaction/coupling with the water environment [54].

Such a dimer/water coupling can lead to a situation analogous to that of atoms interacting with coherent modes of the electromagnetic radiation in quantum optical cavities, namely to the so-called Vacuum-Field Rabi Splitting (VFRS) effect [48]. VFRS appears in both the emission and absorption spectra of atoms [46] in interaction with a coherent mode of electromagnetic radiation in a cavity. For our purposes below, we shall review the phenomenon by restricting ourselves to the absorption spectra case.

Consider a collection of N atoms of characteristic frequency $\omega_0$ inside an electromagnetic cavity. Injecting a pulse of frequency $\Omega$ into the cavity causes a doublet structure (splitting) in the absorption spectrum of the atom-cavity system with peaks at:

$$\Omega = \omega_0 - \Delta/2 \pm (1/2)( \Delta^2 + 4 N \lambda^2)^{1/2} \qquad (9)$$

where $\Delta = \omega_c - \omega_0$ is the detuning of the cavity mode, of frequency $\omega_c$, compared to the atomic frequency. For resonant cavities the splitting occurs with equal weights

$$\Omega = \omega_0 \pm \lambda\sqrt{N} \qquad (10)$$

Notice here the *enhancement* of the effect for multi-atom systems N >> 1. The quantity $2\lambda\sqrt{N}$ is called the 'Rabi frequency' [48]. From the emission-spectrum analysis an estimate of $\lambda$ can be inferred which involves the matrix element, $\underline{p}$, of atomic electric dipole between the energy states of the two-level atom [48]:

$$\lambda = \frac{E_c \underline{d} \cdot \underline{p}}{\hbar} \qquad (11)$$

where $\underline{p}$ is the cavity (radiation) mode polarization, and

$$E_c \sim (\frac{2\pi\hbar\omega_c}{\varepsilon V}) \qquad (12)$$

is the r.m.s. vacuum (electric) field amplitude at the center of a cavity of volume V, and of frequency $\omega_c$, with $\varepsilon$ being the dielectric constant of the medium inside the volume V. In atomic physics the VFRS effect has been confirmed by experiments involving beams of Rydberg atoms resonantly coupled to superconducting cavities [49].

In the analogy between the thin cavity regions near the dimer walls of MTs with electromagnetic cavities, the role of atoms is played by the unpaired two-state electrons of the tubulin dimers [30] oscillating with a frequency (2). To estimate the Rabi coupling between cavity modes and dimer oscillations, one should use (11) for the MT case.

We have used some simplified models for the ordered-water molecules, which yield a frequency of the coherent dipole quanta ('cavity' modes) of order [30]:

$$\omega_c \sim 6 \times 10^{12} \text{Hz} \qquad (13)$$

Notably this is of the same order of magnitude as the characteristic frequency of the dimers (2), implying that the dominant cavity mode and the dimer system are almost in resonance in the model of [30]. Note that this is a feature shared by atomic physics systems in cavities, and thus we can apply the pertinent formalism to our system. Assuming a relative dielectric constant of water with respect to that of vacuum $\varepsilon_o$, $\varepsilon / \varepsilon_o \sim 80$, one obtains from (12) for the case of MT cavities:

$$E_c \sim 10^4 \text{ V/m} \qquad (14)$$

Electric fields of such a magnitude can be provided by the electromagnetic interactions of the MT dimer chains, the latter viewed as giant electric dipoles [53]. This suggests that the coherent modes $\omega_c$, which in our scenario interact with the unpaired electric charges of the dimers and produce the kink solitons along the chains, owe their existence to the (quantized) electromagnetic interactions of the dimers themselves.

The Rabi coupling for the MT case then is estimated from [30] to be of order:

$$\text{Rabi coupling for MT} \equiv \lambda_{MT}$$
$$= \sqrt{N}\lambda_o \sim 3 \times 10^{11} \text{Hz} \qquad (15)$$

which is, on average, an order of magnitude smaller than the characteristic frequency of the dimers (2).

In the above analysis, we have assumed that the system of tubulin dimers interacts with a *single* dipole-quantum coherent mode of the ordered water and hence we ignored dimer-dimer interactions. More complicated cases, involving interactions either among the dimers or of the dimers with more than one radiation quantum (which undoubtedly occur *in vivo*), may affect the above estimate.

The presence of such a coupling between water molecules and dimers leads to quantum coherent solitonic states of the electric dipole quanta on the tubulin dimer walls. To estimate the decoherence time we remark that the main source of dissipation (environment) comes from the imperfect walls of the cavities, which allow leakage of coherent modes and energy. The time scale, $T_r$, over which a cavity-MT dissipates its energy, can be identified in our model with the average life-time $t_L$ of a coherent-dipole quantum state, which has been found to be [30]: $T_r \sim t_L \sim 10^{-4}$ sec. This leads to a first-order-approximation estimate of the quality factor for the MT cavities, $Q_{MT} \sim \omega_c T_r \sim 10^8$. We note, for comparison, that high-quality cavities encountered in Rydberg atom experiments dissipate energy in time scales of $10^{-3}$-$10^{-4}$ sec, and thus have Q's which are comparable to $Q_{MT}$ above. The analysis of [30] yields the following estimate for the collapse time of the kink coherent state of the MT dimers due to dissipation:

$$t_{collapse} \; 10^{-7}\text{-}10^{-6} \text{ sec} \tag{16}$$

This is larger than the time scale (8) required for energy transport across the MT by an average kink soliton in the models of [30, 64]. The result (16), then, implies that quantum physics may be relevant as far as dissipationless energy transfer across the MT is concerned.

Therefore, this specific model is in stark disagreement with the conclusions of Tegmark in [61], i.e. that only classical physics is relevant to the energy and signal transfer in biological matter. Tegmark's conclusions did not take proper account of the possible isolation against environmental interactions, which seems to occur inside certain regions of MTs with appropriate geometry.

### 2.2.3 Ordered Water in Biological Systems

The above scenarios are consistent with independent studies of water in biological matter, which is summarized below.

Recent experimental spectroscopic studies of resonant intermolecular transfer of vibrational energy in liquid water [65] have established that energy is transferred extremely rapidly and along many water molecules before it dissipates. This energy is in the form of OH-stretch excitations and is thought to be mediated by dipole-dipole interactions in addition to a yet unknown mechanism which speeds up the transfer beyond that predicted by the so-called Förster expression for the energy transfer rate between two OH oscillators, k.

$$k = T_1^{-1}\left(\frac{r_o}{r}\right)^6 \tag{17}$$

where $T_1$ is the lifetime of the excited state, r the distance between the oscillators and $r_o$ the Förster radius. The Förster radius, which is a parameter experimentally determined for each material, characterizes the intermolecular energy transfer and has been determined by Woutersen et al. [65] to be $r_o$= 2.1 ±0.05 Å while the typical intermolecular distance (at room temperature) for water is ~2.8 Å. It is evident from these data that the energy transfer in pure water will be fast and yet experimentally it is determined to be even faster than that by one or even two orders of magnitude. Woutersen et al. speculate that this extremely high rate of resonant energy transfer in liquid water may be a consequence of the proximity of the OH groups which causes other, higher-order -uples to also exchange energy. Here, we propose another mechanism to explain the rapidity of the energy transfer, namely kink-soliton propagation. This is based on the phenomenological realization that it is exactly this kind of energy transfer that one would expect to see experimentally as a result of the existence of kink-solitons. Such a mechanism, regardless of exact origin, is ideal for loss-free energy transfer between OH groups located on either different biomolecules or along extended biological structures such as MTs which would be covered (inside and out) with water. Note also that this predicts that OH groups in hydrophobic environments would be able to remain in a vibrationally excited state longer than OH groups in hydrophilic environment lending credence to our working assumption that the electrons inside the hydrophobic pockets of the tubulin molecules are sufficiently isolated from thermal noise.

It is quite possible that such solitonic states in water may not be quantum in origin in the case of microtubules. The 25 nm diameter of the MT is too big a region to allow for quantum effects to be sustained throughout, as we discussed above. Such solitons may be nothing other than the ones conjectured in [66], which may be responsible for the optical transparency of the water interior of MTs. Such classical solitons in the bulk of the water interior may co-exist with the quantum coherent states on the dimer walls [30].

### 2.2.4 The Relevance of Superconductivity to Quantum Coherence in Biological Matter.

As we have already seen, some of the principal objections to the quantum consciousness idea are that: (1) thermal fluctuations and interactions with the environment will destroy quantum coherence over distances greater than a few nanometers at biological temperatures, and (2) the total mass of the particles is too large. However, the discovery of high temperature superconductivity [67] establishes the existence of large-scale quantum coherence at temperatures within a factor of three of biological temperatures. MRI magnets contain hundreds of miles of superconducting wire and routinely carry a persistent current. There is no distance limit – the macroscopic wave function of the superfluid condensate of electron pairs, or Cooper pairs, in a sufficiently long cable could maintain its quantum phase coherence for many thousands of miles[2]. Moreover, there is no limit to the total mass of the electrons participating in the superfluid state. The condensate is "protected" from thermal fluctuations by the BCS energy gap at the Fermi surface, and the term "quantum protectorate" has recently been coined [68] to describe this and related many-body systems.

A remarkable phenomenon displayed by superconductors is that of flux quantization [69], in which the magnetic flux $\Phi$ passing through a superconducting ring is quantized in multiples of $\Phi_0 = h/2e$ (in SI units, or $hc/2e$ in cgs units). This is because the Cooper pairs are condensed into the same macroscopic wavefunction, or condensate, whose phase must change by a multiple of $2\pi$ as one traces a closed path around the ring. One consequence is that the angular momentum of *each* Cooper pair is quantized in multiples of $\hbar$. Amazingly, the *total* angular momentum of the electron superfluid is also quantized, but in multiples of $N_p\hbar$, where $N_p$ is the number of pairs. The magnetic flux is then proportional to the total angular momentum divided by the total circulating charge, i.e. $\Phi_0 = 2\pi N_p\hbar/\{N_p(2e)\} = h/2e$. It is as though the effective quantum of angular momentum (or of action) and that of charge had been scaled up to macroscopic quantities. A similar phenomenon occurs in superfluid helium, where the circulation, defined as the integral of the velocity around a closed loop, is quantized.

Condensed matter systems providing evidence for macroscopic quantum effects at *biological* temperatures include charge and spin density waves in quasi-1-D conductors[70]. A Peierls gap opens up at the Fermi surface, thus creating a "quantum protectorate" below the transition temperature. For example, the quasi-1-D compound $NbS_3$ forms a density wave (DW) at 340 K or 67ºC, 30ºC higher than the human body temperature, and the transition temperatures are even higher in some quasi-2-D compounds. A DW pinned by impurities can collectively transport an electric current when an applied electric field exceeds a threshold value. A simplified model represents the DW as an elastic string in a washboard, or sine-Gordon, pinning potential. In the classical picture, a sufficiently large field tilts the washboard potential enough for the DW to "slide." (Real pinning potentials are disordered, but coherent voltage oscillations seen in high quality crystals suggest that the simple sine-Gordon picture captures much of the essential physics.)

---

[2] What is generally referred o as the "coherence length" ξ of a superconductor is actually the distance over which the amplitude of the order parameter can vary significantly. The phase coherence length is essentially infinite. Otherise, flux quantization could not occur in a persistent current mode.

A number of experiments indicate that nucleation of soliton-antisoliton (SŜ) domain wall pairs [71], via quantum tunneling [72, 73], or "decay of the false vacuum" [74] occurs for fields far below the classical depinning threshold in weakly pinned DWs. In particular, NMR experiments[75, 76] and an observed bias-independent ac response below threshold [77] show that the DW phase is displaced by only a fraction of its classically predicted value as the applied field approaches threshold in some materials. These experiments are explained nicely by the quantum picture, in which the threshold for SŜ nucleation can be much smaller than the classical depinning field, and is a Coulomb blockade effect due to S-Ŝ electrostatic interactions [78, 79]. Moreover, *rf* linear response and mixing experiments [80, 81] show remarkable agreement with the predictions of photon-assisted tunneling theory and, indirectly at least, give a scaling between voltage and frequency that is consistent with the ratio $\hbar/e$ [82]. Aharonov-Bohm type oscillations in the magnetoconductance of DWs with columnar defects [83] and "soliton-tunneling transistor" experiments [84] provide further evidence for quantum effects, the latter showing evidence for macroscopically charged domain walls. Finally, the shapes of the current-field (*I-E*) characteristics of weakly pinned DWs show nearly precise agreement [82] with the form $I \sim [E - E_T] \exp[-E_0/E]$ predicted by the soliton tunneling model, where $E_T$ is the Coulomb blockade threshold field for SŜ pair creation [85] and $E_0$ depends on the washboard pinning potential.

Some theoretical arguments for how topological soliton domain walls might suppress decoherence are as follows. The tunneling probability amplitude is governed by the action *density* rather than the total action for nucleating SŜ domain wall pairs, and is thus independent of the transverse dimensions of the domain walls (or "branes") [86]. Even infinitely large domain walls would thus have a finite probability of nucleating via quantum tunneling. However, thermal fluctuations and decoherence effects will be suppressed by the topological stability and overall large mass-energy of the domain walls, thus allowing coherent quantum effects to dominate even at high temperatures. Perhaps biological systems use similar mechanisms to suppress decoherence, and thereby enable the utilization of large-scale quantum coherence. These systems are of potential relevance to biology because: (1) quasi-1-D structures, including microtubules, actin filaments, the DNA helix, etc., are pervasive in biological systems, (2) the transition temperatures can be comparable to or higher than biological temperatures, (3) the mechanisms by which decoherence is suppressed may be similar in biological systems, and (4) soliton models have been proposed for several biological systems, including DNA and microtubules.

A type of learned behavior, known as the pulse-duration memory effect, is another phenomenon especially pertinent to the QCI. It has been found that, by applying a series of rectangular current pulses of identical width, one can "train" a DW to remember the widths of the previous pulses, such that a voltage oscillation minimum produced by the DW response coincides with the end the next applied pulse, apparently defying causality [87]. What is especially intriguing is that just a few pulses (in some cases only one!) are required for the DW to "learn," whereas classical simulations predict that many (often hundreds) of pulses should be needed. This suggests that a driven DW may operate as a rudimentary quantum computer operating at relatively high temperatures. The key may be the redundancy provided by soliton domain walls, each of which contains identical kinks along many parallel chains. The price paid is the increased size of each elementary unit of information, but the gain is that the domain walls are protected from decoherence and thermal fluctuations. Perhaps biological systems also use some degree of redundancy and/or topological objects to suppress decoherence. Clearly, further experiments, as well as more detailed theoretical work on possible mechanisms, are strongly warranted.

### 2.2.5 Error Correction and Quantum Entanglement of a Macroscopic System

As we have seen above, under appropriate environmental isolation, it is theoretically conceivable to obtain quantum coherence of macroscopic populations of tubulin dimers in microtubule systems, which can be sustained for long enough times so that dissipationless energy and signal (information) transfer can occur in a cell.

We would now like to discuss the feasibility of the above, admittedly speculative, ideas by making a brief report on recent progress made in experimentally demonstrating macroscopic quantum entanglement at room temperature in atomic physics.

In a recent article [50] Julsgaard et al. describe the macroscopic entanglement of two samples of Cs atoms at room temperature. The entangling mechanism (see Section 5 for more on entanglement) is a pulsed laser beam and although the atoms are far from cold or isolated from the environment, partial entanglement of bulk spin is unambiguously demonstrated for $10^{12}$ atoms for ~0.5 ms. The system's resilience to decoherence is in fact *facilitated* by the existence of a large number of atoms as even though atoms lose the proper spin orientation continuously, the bulk entanglement is not immediately lost. Quantum Informatics, the science that deals with ways to encode, store and retrieve information written in qubits has to offer an alternative way of interpreting the surprising resilience of the Cs atoms by implicating redundancy. Simply stated, information can be stored in such a way that the logical (qu)bits correspond to many physical (qu)bits and thus are resistant to corruption of content. Yet another way of looking at this is given in the work by Kielpinski et al. [88] a decoherence-free quantum memory of one qubit was built by encoding the qubit into the "decoherence-free subspace" (DFS) of a pair of trapped Beryllium $^9Be^+$ ions. They achieved this by exploiting a "safe-from-noise-area" of the Hilbert space for a superposition of two basis states for the ions, thus encoding the qubit in the superposition rather than one of the basis states. By doing this they achieved decoherence times on average an order of magnitude longer that regular. Both of the above works show that it is possible to use DFS, error correction and high redundancy to both store information and to keep superpositions and entanglements intact for biologically relevant times in macroscopic systems at high (room) temperature. Thus, it may not be entirely inappropriate to imagine that in biological *in vivo* regimes, one has, under certain circumstances such as specified above, entanglement of tubulin/MT arrangements.

## 2.3    Implications for Cell Function

The above raises the question of how such phenomena can affect the functioning of cells. In other words, would the existence of such coherent states and the emergence of quantum mechanical entanglement be somehow useful or beneficial to biological function? Is it then reasonable to propose that in certain cases, natural selection may have favored molecules and cellular structures that exhibited and sustained such phenomena? If we accept the notion that according to the laws of quantum physics certain macroscopic arrangements of atoms will exhibit such effects, is it not reasonable then to expect that biomolecules and (by extension) cellular structures and whole cells have 'found' a use for such phenomena and have evolved to incorporate them? We stress that at a given instant in time, the different microtubule coherent states participating in a specific bulk entanglement would be almost identical due to the fact that they are related/triggered by a specific "external agent" (e.g. the passing of a specific train of action potentials in the case of a neural cell). This is of importance since it increases the system's resilience to decoherence (by entangling a large number of nearly identical states), in addition to facilitating "sharp decision making" (i.e. rapid choice among a vast number of very similar states) as explained in [13] which is presumably a trait favored by natural selection. More and more, physicists are exploring the quantum mechanics behind such important biological processes as enzyme action [10] or photosynthesis [11]. Below, we digress to investigate one possible way teleportation of coherent quantum states across and between cells can play a biologically-relevant role as the mechanism behind some of the information storage, retrieval and processing done by cells.

### 2.3.1 Biological Quantum Teleportation

We define teleportation as the complete transfer of the coherent state of an MT *without any direct transfer of mass or energy*. This means that the 'receiver' MT finds itself in an identical state to the 'sender' MT. We will demonstrate the way in which, given the possibility for entangled states, teleportation between microtubule A and microtubule C can occur. The use of pure state vectors, |Ψ>, to describe the coherent states along a MT arrangement is justifiable since they do not obey the ordinary Schrödinger evolution equation. Instead, they obey the stochastic equations of open systems, of the form discussed in [89]. Nowhere in the proof of teleportation below are the precise forms of the evolution equations used. As argued in [89], by using appropriate stochastic (Langevin type) equations one may recover, for instance, the standard Lindblad form of evolution equations for the corresponding density matrices $\rho = Tr_M|\Psi><\Psi|$, where M is an appropriate subset of environmental degrees of freedom, non-accessible to the observer.

A coherent state in microtubule A (referred to as simply A and designated as |Ψ(A)>) of the (collective) dipole moment(s) being in either of the two classically allowable states with probability amplitude $\omega_0$ and $\omega_1$ can be written as:

$$|\Psi(A)> = \omega_0|0> + \omega_1|1> \tag{18}$$

**Step 1:** The cell finds itself with microtubule B and microtubule C -which can be parallel or collinear- in an entangled state written as:

$$|\Psi(B,C)> = \quad (1/(\sqrt{2})) \ |1_B,0_C> + |0_B,1_C> \tag{19}$$

The combined state of A,B,C can be written as:

$$|\Psi(A,B,C)> = |\Psi(A)> \otimes |\Psi(B,C)> \tag{20}$$

which upon expanding the outer product $\otimes$ can be written as:

$$\left|\Psi(A,B,C)\right\rangle = \frac{1}{\sqrt{2}}\left\{\omega_o\left(\left|0_A,1_B,0_C\right\rangle + \left|0_A,0_B,1_C\right\rangle\right) + \omega_1\left(\left|1_A,1_B,0_C\right\rangle + \left|1_A,0_B,1_C\right\rangle\right)\right\} \tag{21}$$

We can also express the combined state |Ψ(A,B,C)> in a different basis, known as the "Bell basis". Instead of |0> and |1>, the basis vectors will now be,

$$\left|\Psi^{\pm}(A,B)\right\rangle = \frac{1}{\sqrt{2}}\left(\left|0_A,1_B\right\rangle \pm \left|1_A,0_B\right\rangle\right) \tag{22}$$

and

$$\left|\Phi^{\pm}(A,B)\right\rangle = \frac{1}{\sqrt{2}}\left(\left|0_A,0_B\right\rangle \pm \left|1_A,1_B\right\rangle\right) \tag{23}$$

In this new basis, our state of the three microtubules |Ψ(A,B,C)> is written as:

$$\left| \Psi(A,B,C) \right\rangle = \frac{1}{2} \Big( \left| \Psi^+(A,B) \right\rangle \otimes \big( \omega_o \left| 0_C \right\rangle + \omega_1 \left| 1_C \right\rangle \big) \Big) +$$
$$\left| \Phi^+(A,B) \right\rangle \otimes \big( \omega_o \left| 1_C \right\rangle + \omega_1 \left| 0_C \right\rangle \big) + \left| \Psi^-(A,B) \right\rangle \otimes \big( \omega_o \left| 0_C \right\rangle + \omega_1 \left| 1_C \right\rangle \big) + \qquad (24)$$
$$\left| \Phi^-(A,B) \right\rangle \otimes \big( \omega_o \left| 1_C \right\rangle + \omega_1 \left| 0_C \right\rangle \big)$$

This concludes the first step of teleporting the state of MT A to MT C.

**Step 2:**     Notice that so far, MT A has not interacted with the environment or cell, i.e. the coherent state of A which we designated as |Ψ(A)>=ω$_o$|0$_C$>+ω$_1$|1> has not been touched. Now the part of the cell containing A and B (the "sender part") makes a "measurement" -which in our case can be an electromagnetic interaction with a passing action potential or the binding of a MAP molecule. If this measurement or forced collapse is done in the Bell basis, on |Ψ$^\pm$(A,B)> it will project the state in MT C (!) to:

$$\left| \Psi^\pm(C) \right\rangle = \left\langle \Psi^\pm(A,B) \middle| \Psi(A,B,C) \right\rangle = \omega_o \left| 0_C \right\rangle \pm \omega_1 \left| 1_C \right\rangle \qquad (25)$$

similarly

$$\left| \Phi^\pm(A,B) \right\rangle \rightarrow \left| \Phi^\pm(C) \right\rangle = \omega_0 \left| 1_C \right\rangle \pm \omega_1 \left| 0_C \right\rangle \qquad (26)$$

This effectively concludes the teleportation of the state of MT A to MT C with one caveat.

**Step 3:** There is a probabilistic nature to this process, which means that MT C may receive the exact copy of the state of MT A i.e. |Ψ$^+$(C)> or it may receive a state which is a unitary transformation away from the original |Ψ(A)> (one of the other three possibilities: |Ψ$^-$> or |Φ$^\pm$>. MT C can reproduce the state of MT A if there is a 'hardwired' condition so that when MT C receives |Ψ$^+$(C)> it does nothing further, yet if it receives one of the other three, it performs the correct unitary transformation to obtain the correct state from A. This 'hardwired' behavior can be implemented through the use of codes, not unlike the Koruga bioinformation [37] code that MTs seem to follow. In principle, the exact state correspondence may not even have any significance and instead the information could be encoded in the frequency of transferred states, similar to information being encoded in the frequency of action potentials and not their shapes.

Teleportation is an entirely non-classical phenomenon and has been experimentally demonstrated in matter and light states and combinations of those (see Section 5). Currently, the scientific consensus is that teleportation is impossible without entanglement although this may change in the future. Biological teleportation as described above, can be imagined as the basis of intra- and inter- cellular correlation which leads to yoked function (e.g. intracellularly during translation and intercellularly during yoked neuron firing). Experiments to check for such teleportation of states can be designed based on the Surface Plasmon Resonance (SPR) principle [90] as applied to sheets of polymerized tubulin immobilized on a metal film (Sections 4 and 5). A graphical representation of biological quantum teleportation of dipole moment states is presented in Fig. 2.

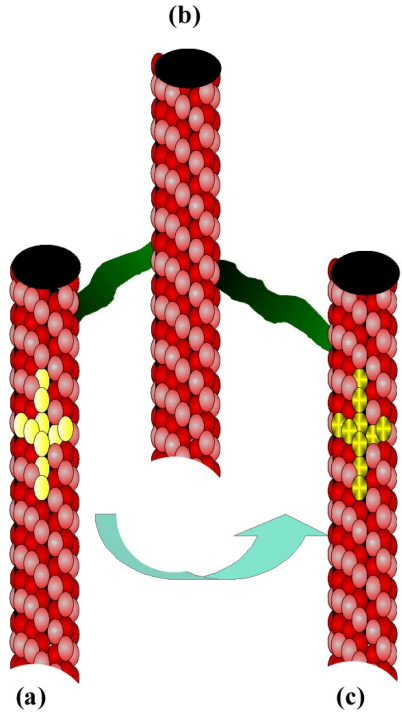

**Fig. 2. Schematic of Quantum Teleportation of Dipole States**. (Taken from [16]) MT **(a)** sends its state (represented by the yellow cross) to MT **(c)** without any transfer or mass or energy. Both MT (a) and MT (c) are entangled with MT **(b)**. Entanglement represented by the presence of connecting MAPs (green).

### 2.3.2 Information Processing by Biopolymers and the Guitar String Model (GSM)

In the quantum-mechanical scenario for MT dynamics discussed above, as suggested in [30], a quantum-hologram picture for information processing of MT networks emerges. Further, the existence of solitonic quantum-coherent states along the MT dimer walls implies a role for these biological entities as logic gates [14]. Consider, for instance, a node (junction) of three MTs connected by microtubule associated proteins (MAPs) see Fig. 3. The quantum nature of the coherent states makes the junction interaction probabilistic. Therefore at tube junctions one is facing a Probabilistic Boolean Interaction. The probability for having a solitonic coherent state in a MT branch does depend on its geometric characteristics (such as length). By modulating the length of the tubes and the binding sites of the MAPs a bias can be introduced between bit states which can affect the probabilistic final outcomes. This has obvious implications for information processing by MT networks.

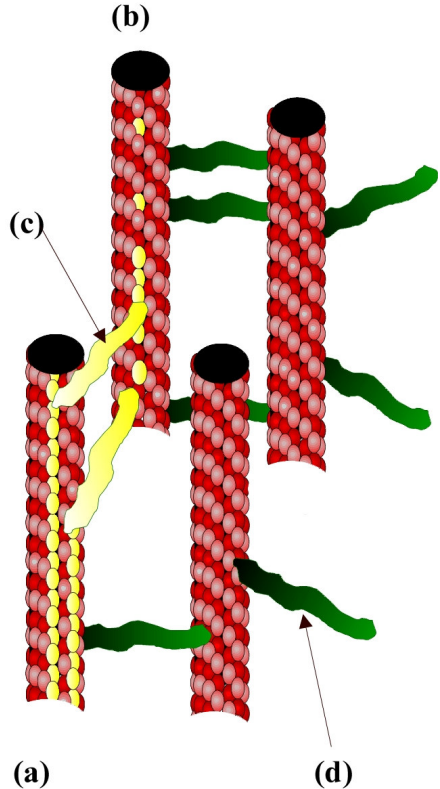

**Fig. 3. An XOR (Exclusive-OR) Logic Gate**. (Taken from [16]) "0" ("1") is represented by absence (presence) of solitonic kink wave of dipole moment flips indicated in yellow. **(a)** Input MT, **(b)** Output MT, **(c)** A MAP transmitting a solitonic kink wave, **(d)** a 'quiet' MAP. MT (a) has two solitons traveling, encountering two MAPs that transmit both solitons to MT (b). In this hypothetical scenario, the solitons arrive out of phase at MT (b) and cancel each other out. The truth table for XOR reads: 0,0→ 0; 0,1→1; 1,0→1, 1,1→0 and can be realized by an MT arrangement if the MAPs are arranged such that each can transmit a soliton independently but if they both transmit the solitons cancel out, i.e. the two MAPs must be an odd number of dimers apart on MT(a).

Such a binary information system can then provide the basic substrate for quantum information processing inside a (not exclusively neural) cell. In a typical MT network, there may be about $10^{12}$ tubulin dimers. Although such a number is large, as discussed earlier there may be subtle "shielding" mechanisms at play. The above scenario is not necessarily *quantum* in nature. An essentially identical argument can be made for information processing via waves of dipole flips or just momentum transfer as a result of propagating conformational changes.

This suggests an obvious model for encoding information in a network of MTs which we shall call the "Guitar String Model" to emphasize the analogy to the way six guitar strings (the MTs) can by clamped by four fingers (MAPs) at different nodes to generate hundreds of different chords (engrams). If propagating dipole moment flips are indeed carrying signals inside the cell then the nodes in the network can affect this propagation in a large variety of ways. A limited set of MTs with a limited set of MAP binding sites can have a very large set of engrams. This also suggests a way for new memories to form and old to be erased by simply changing the distribution of MAPs.

### 2.4 Conclusions

If it is experimentally confirmed that treating MTs as QED cavities is a fair approximation to their behavior, one can propose that nature has provided us with the necessary structures (microtubules) to operate as the basic substrate for quantum computation either in vivo, e.g. intracellularly, or in vitro, e.g. in fabricated bioqubit circuits. Such a development would pave the way to construct quantum computers by using microtubules as building blocks, in much the same way as QED cavities in quantum optics are currently being used in successful attempts at implementing qubits and gates [59]. Detecting quantum behavior at this level would undoubtedly advance attempts at implicating quantum physics in consciousness.

**SECTION 3:   TAU ACCUMULATION IN *DROSOPHILA* MUSHROOM BODY
NEURONS RESULTS IN MEMORY IMPAIRMENT**

**3.1     Introduction**

In this Section we summarize and attempt a "physicist friendly" account of the neurobiological results obtained by our group and published in [9].

In order to test some of the predictions of the models discussed in Sections 1 and 2, an in-vivo neurobiological behavioral study was undertaken. The goal was to experimentally investigate whether memory is affected by perturbations in the microtubular (MT) cytoskeleton. Associative olfactory learning and memory were the types of memory accessible to us with the transgenic *Drosophila* fruitfly behavior analysis system. We tried disturbing the fly MTs as little as possible, avoiding perturbing the cytoskeleton by formation of such large protein aggregates as neurofibrillary tangles (NFTs) that could effectively 'strangle' the neuron disrupting or even stopping intracellular (axonal) transport. In addition, NFTs and/or amyloid or senile plaques (APs or SPs) have been unequivocally shown to contribute to neurodegeneration and eventual neuronal death and it is reasonable to expect a dying neuron to dysfunction, regardless of the state of its MTs. We also avoided causing any developmental problems by selecting gene promoters (drivers) with appropriate temporal activity.

Since this is a contribution addressed mostly to physicists an effort has been made to explain potentially unfamiliar biological terms and procedures. Following established standards in genetics, small case italics such as *tau* indicate the *gene* that codes for the protein TAU indicated in capitals. Strains or lines of transgenic animals (animals that have been genetically manipulated and contain extra genes) are named somewhat arbitrarily so here we use *b-, h-,* or *d-,* indicating bovine- human- or *Drosophila-* (native) derived- genes while a few letters identifying the source.

Microtubule associated protein (MAP) TAU has long being implicated in the encoding of human memory and it has been shown that mutations in the human NC-17 *tau* gene are one of the causes of Alzheimer's Disease [91-93]. For this reason, NFT and SP/AP formation have been the main focus of studies of tauopathies in animal models. For instance, transgenic mice with Fronto-Temporal Dementias with Parkinsonism (FTDP 17) mutations develop NFTs and neurodegeneration accompanied by motor deficits [94, 95] Expression of human wild type and FTDP-17-linked mutations in *Drosophila* results in age-dependent neurodegeneration without NFTs [96] except when wild-type TAU was phosphorylated by overexpressed *Drosophila* glycogen synthase kinase-3 [97]. Mice carrying mutated *tau, presenilin 1* and *alpha-beta Peptide Precursor (APP)* transgenes show synaptic dysfunction before the development of NFTs or amyloid plaques. From these and other studies it seems that tauopathy-caused deficits in memory appear even without NFTs or SPs/APs although frequently, at least NFTs do eventually appear in the late stages of the disease.

For NFTs to form there must be a situation of elevated TAU accumulation (in a non-filamentous or 'pre-tangle' state) in the affected neurons. Such a condition has been suggested as the underlying cause of pre-neurodegeneration cognitive symptoms such as memory loss [94, 98] and our research experimentally addresses the question of the effect of elevated pre-tangle state TAU in *Drosophila* mushroom bodies and we propose a connection between the observed effects and theoretical models of cytoskeletal function.

**3.2     *Drosophila***

The *Drosophila Melanogaster* fruit fly has long been a favorite of experimental behavioral neurobiologists for numerous reasons including its relatively simple genetic makeup and quick generation time, powerful classical and molecular genetics and the animal's ability to learn and remember a variety of tasks.

To illustrate our approach and choice of *Drosophila* more fully, our initial experimental design will be briefly described here. *Drosophila* was selected as the ideal system for investigating cytoskeletal involvement in learning and memory because we were to attempt to track an intraneuronal *redistribution* of MAP-2 and/or MAP TAU as a result of conditioning. This is a prediction of the GSM described in Section 2. In order to track a redistribution of MAPs *inside* neurons one must be able to differentiate between the various parts of the neuron such as the dendrites, axons, axonal projections and somata. In humans and other mammals, the neuronal organization is such that multiple neurons and neuronal types are involved in a given process forming an extensive complex network of axons and dendrites. As a result, it is particularly difficult to locate individual neurons' specific parts and stain selectively to track changes in distribution of a particular protein. In *Drosophila* on the other hand, the neuronal organization is such that differentiation of subneural parts is facilitated. For instance, neurons belonging to mushroom bodies (MBs are prominent structures in the *Drosophila* brain essential for olfactory learning and memory) represent a highly ordered, tightly and sequentially packed neuronal system where axonal projections (i.e. synaptic fields), dendrites and somata are macroscopically (order μm) separated in ordered fiber bundles, see Fig. 4.

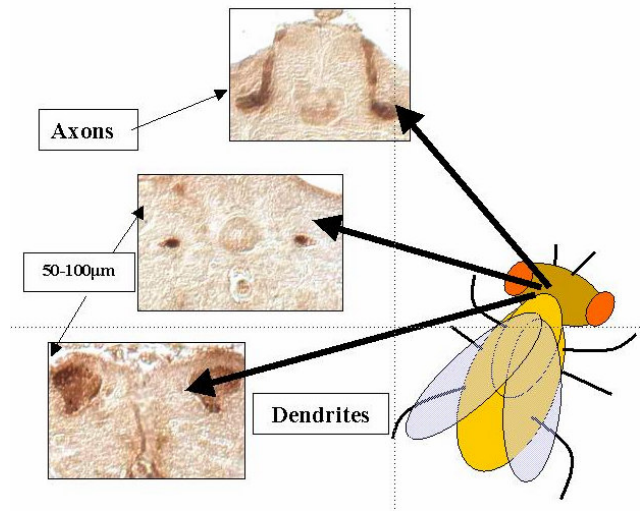

**Fig. 4. Fly Mushroom bodies (MBs).** Fly mushroom bodies are shown in paraffin frontal sections ~5μm thick, stained for LEO, a MB specific protein.

This provides a strong advantage for analysis of the results of expression of microtubule associated proteins in specific neurons (e.g. those associated with a specific type of memory) but also *within* different parts of such neurons. For instance, a bulk redistribution of a certain MAP from the axons to the dendrites of the MB, presumably as a result of memory formation can, in principle, be tracked. This is in fact a prediction of the GSM for memory encoding since if the MAPs play the role of fingers on the guitar fret board and the various chords correspond to encoded information, acquisition of new information and memory would result in a redistribution of MAPs. Unfortunately, our preliminary experiments utilizing directed expression of chicken MAP-2 in MBs showed that either the resolution offered by existent anti-MAP antibodies was insufficient to decipher appreciable MAP redistribution and/or no such redistribution took place as a result of learning. The latter would be inconsistent with results obtained in rodents [8] that suggest a redistribution of MAP-2 resulting in accumulation in dendrites as a result of learning an auditory associative task.

We therefore shifted our approach to determining the impact of MAP TAU *overexpression* on the ability of the animals to learn and retain memories. Although this is not as direct a test of the GSM it does provide a solid link between the microtubular cytoskeleton and memory retrieval and stability as will be argued in this Section.

### 3.3 Genetic Engineering

We induced the expression of vertebrate (human and bovine) *tau* genes, producing TAU protein in *specific* tissues and at specific *times* in *Drosophila* using the method of directed gene expression.

### 3.3.1 Directed Expression

Directed gene expression rests on the principle of obtaining two genetically manipulated (transgenic) lines, the first of which contains the gene to be expressed, fused to and under the direction of an *upstream activating sequence* (UAS). This UAS promoter is activated by the presence of its unique, selective and specific activator protein GAL4.

To generate transgenic lines expressing GAL4 in a cell or tissue specific pattern, the GAL4 gene is inserted randomly into the fly's genome, thus driven in its expression from various genomic enhancers. A GAL4 target gene (UAS-tau) will remain silent in the absence of GAL4. To activate the target gene, the flies carrying the UAS-tau are crossed to flies expressing GAL4 at specific tissues and at specific times in the animal's development (see Fig.5). To eliminate potential complications arising from expression of TAU in the embryonic and developing nervous system, we selected strains expressing GAL4 in late pupal and adult mushroom body neurons [99] only by utilizing the MB drivers c492 and c772.

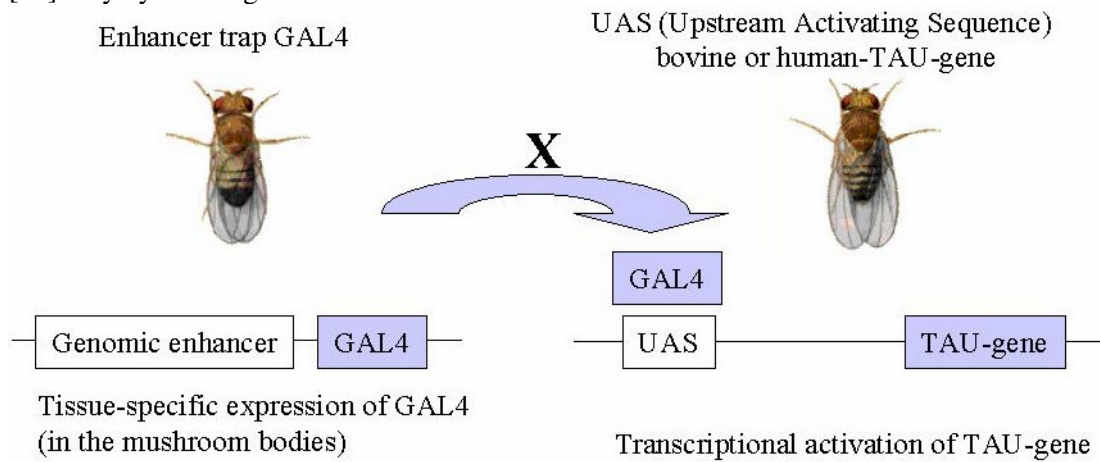

**Fig. 5. Upstream Activating Sequence and Target Gene.**

### 3.3.4 Coimmunoprecipitation

Note that past studies have shown that mere accumulation of non-*Drosophila* proteins such as β-galactosidase [100, 101], and GAL4, or *Drosophila* proteins [102, 103] that are *not* MT specific in MB neurons, do *not* cause any behavioral deficits. Therefore, once TAU is shown to bind to MB MTs, any effects of TAU accumulation in MBs can be taken as specific to TAU. Also note that mere presence of two proteins in the same tissue does not necessarily mean they are bound together so although we expected h- and b- TAU to bind to MTs we performed a coimmunoprecipitation study. The *Drosophila* protein (dTAU), contains four putative tubulin binding repeats [104]. They exhibit 42% and 46% identity (62% and 66% similarity)[*] with the respective sequence of bTAU and hTAU [104-106]. To determine whether this sequence conservation among vertebrate and *Drosophila* TAU signals also a functional conservation (i.e. if all types retain their MT binding sites intact), anti-tubulin antibodies were used in immunoprecipitation experiments from head lysates of *btau*-expressing animals and controls. We found that bTAU co-immunoprecipitates with *Drosophila* tubulin, indicating that the vertebrate protein is capable of binding *Drosophila* microtubules. Figures and extensive details are presented in [9].

There is one obvious problem in assuming that we have just substituted 'more of the same' in the fly's mushroom bodies because dTAU lacks the amino-terminal extension of vertebrate TAU and this suggests that the conformation of vertebrate TAU will be somewhat different from the *Drosophila* protein despite its microtubule-binding ability. This however does not affect the conclusions of this study as will be illustrated later.

Collectively, the ability to bind *Drosophila* tubulin in head lysates and its preferential accumulation within the mushroom bodies indicate that bTAU, and by virtue of its high degree of identity hTAU, bind to the microtubular cytoskeleton within these neurons. Therefore, in the mushroom body neurons of the three *tau* transgenics investigated, the microtubular cytoskeleton is likely burdened with excess TAU.

All strains were normalized to an isogenic (i.e. genetically identical) $w^{1118}$ strain[*]. To obtain flies for behavioral analyses, c772 and c492 homozygotes were crossed to UAS-*btau*, UAS-*htauwt1*, UAS-*dtau 4* and UAS-*dtau1* homozygotes (see secparate section on *dtau* below) and the progeny was collected and tested 3-5 days after emergence. Similarly, the UAS-*btau*, UAS-*htauwt1*, UAS-*dtau 4* and UAS-*dtau1* homozygotes were crossed to $w^{1118}$, the line not containing any drivers (and thus one does not expect to see any extra tau expression) to obtain heterozygotes used as controls.

In addition to the pattern of expression of a gene, it is important to also quantify the amount of protein that is being created. To investigate the relative level of TAU accumulation within adult fly heads we performed semi-quantitative Western analyses (see [9] for details) We determined that bovine TAU was present in head lysates of animals that had the *tau* transgenes and the MB drivers, but not in parental strains as was expected. The level of bTAU protein did not appear to change significantly over a three-week period, indicated by densitometric quantification of results from three independent experiments. Similarly, the level of hTAU appeared relatively constant (data not shown).

Although it would be desirable to know exactly how much more TAU than normal is present, the techniques we used for quantifying the presence of b and h protein relied on mono- or polyclonal antibody binding and densitometry and thus are inherently difficult to normalize.

In principle, it would be possible to refine these findings with such elaborate methods as ion trapping and matrix-assisted laser directed ionization and time- of-flight spectroscopy. Even though we did not attempt such high degree of quantification, we are confident that the transgenic animals did express a significantly higher level of TAU in their mushroom bodies and this is sufficient for the scope of our study.

---

[*] Identity is defined as absolute conservation of the amino acid sequence between two proteins while similarity is conservation of type (e.g. exchanging one acidic amino acid for another acidic preserves similarity)

[*]Isogenic lines are strains of identical genetic background. $w^{1118}$ was chosen to represent the wild type genotype. The transgene of interest was bound to red-eye phenotype and the trangenic flies were crossed to $w^{1118}$ for seven generations (keeping only the red eyes) thus normalizing the genetic background and avoiding contamination.

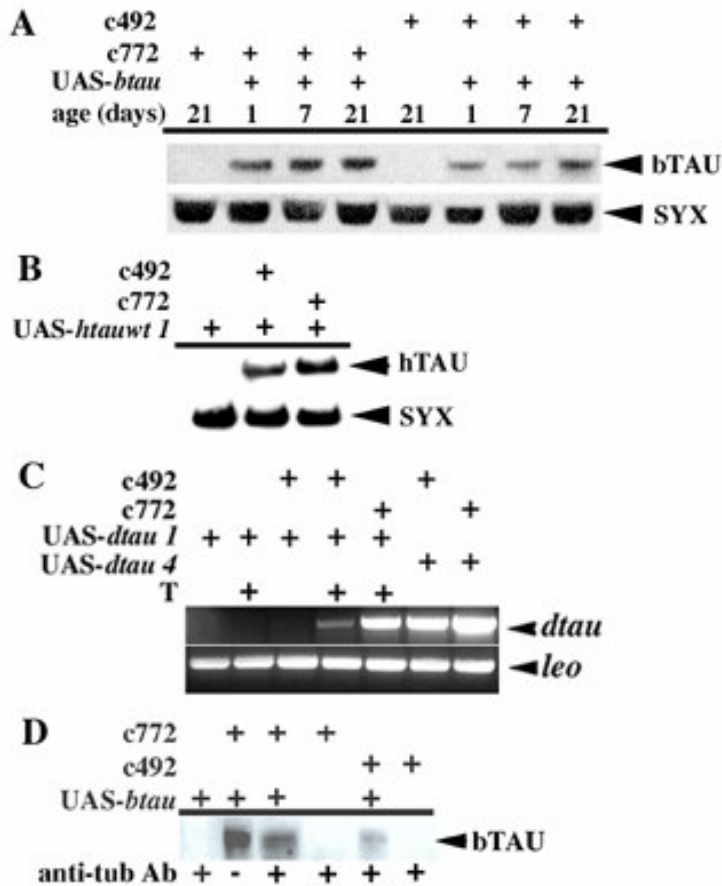

**Fig. 6** (Taken from [9])To determine whether the vertebrate (bovine) TAU binds onto *Drosophila* microtubules, extracts were prepared from the brains of flies that express UAS-*btau* under the c772 (lanes 2 and 3) and the c492 (lane 5 drivers) under mild conditions that would not disrupt TAU/tubulin complexes. A monoclonal antibody against tubulin was added to the mix and then precipitated using standard techniques (co-immunoprecipitation). The resulting complexes were resolved in a denaturing poly-acrylamide gel blotted onto a nylon membrane and the presence of bovine TAU was investigated using a monoclonal antibody. Lane 1 does not contain bTAU because the extracts were prepared from heads of animals that contained only the btau transgene without the driver and therefore do not accumulate the vertebrate protein. Lane 2 represents extracts from bTAU accumulating animals under the c772 driver that were not subjected to co-immunoprecipitation. Therefore the characteristic bTAU band serves as a marker of the size and the amount of total bTAU in the extract. Lane 3 represents that products of co-immunoprecipitation from an equal amount of protein as in lane 2. Detection of a bTAU specific band demonstrates that most of the vertebrate protein (compare the intensities the band in lane 2 indicating the total amount of bTAU in the head extract with the intensity of the band in lane 3 indicating the amount of bTAU that is complexed with Drosophila tubulin in the extract) exists in a complex with the *Drosophila* tubulin. Similar results were obtained with the c492 driver (lane 5). Notably, these bTAU specific bands were absent in the negative controls (extracts from animals not expressing bTAU because they do not carry the UAS-btau transgene, but only the drivers (lanes 4 and 6).

## 3.4 Conditioning

*Drosophila* fruit flies are naturally attracted or repulsed with a variety of affinities by different odors. We followed two standard negatively reinforced associative learning paradigms that essentially generalize the Pavlovian conditioning protocol by coupling aversive odors as conditioned stimuli (CS+ and CS-), and electric shock as the unconditioned stimulus (US). This way, olfactory cues are coupled with electric shock to condition the flies to avoid the odorant associated with the negative reinforcer. These conditioning protocols for *Drosophila* were initially developed by Tully & Quinn [107] and modified by E.M.C. Skoulakis et al. [101, 108]. We used two aversive odorants: 3-Octanol (OCT) and Benzaldehyde (BNZ). The conditioning apparatus consists of a training chamber and a selection maze (see Fig. 7). The maze is normalized by adjusting the concentration of odorants. Once normalized, both wild type (control) and transgenic naïve (i.e. untrained) flies choose to enter one of two identical tubes smelling of OCT and BNZ respectively, with a probability of 50% (as they avoid both odors equally). Because the earliest possible time that we can test the animals past the CS+ and US presentation is 180-200 seconds, our measurements cannot differentiate between "acquisition" and "3-minute memory". This earliest performance assessment will be referred to as "learning".

Conditioning of the flies in the LONG training protocol takes place as follows. A batch of wild type, naïve flies (numbering between 50 and 60) are collected under light anesthesia (using $CO_2$) and 12-24 hours later are left in the dark for one to two hours. The entire conditioning procedure takes place in a temperature- and humidity-controlled darkroom under red illumination in order to isolate the effects of olfactory stimulation from visual stimulation (flies have been shown to react least to red light). Once the flies have been acclimated to the darkroom, half are inserted into conditioning chamber A. The cylindrical wall of the chamber is covered by a grid of two interspersed conducting electrodes spaced such that at least two of the fly's six legs must be in contact with two opposite voltage electrodes. The diameter of the chamber is chosen such as to prevent flies from hovering midair. This way, the fly necessarily completes the circuit making it the path of least resistance. The electrodes are electrified by a signal generator set to 92.0V. The flies receive eleven electric shocks once every five seconds. During this time, the chamber is filled with air containing OCT. The flies are given 30 seconds to rest while the air is being cleared of odorants and are then given the opposite (control) odorant (in this case BNZ) for another minute in the absence of electrical shocks. A rest period of 30 seconds follows after which the flies are tested for acquisition of memory by being inserted into the selection maze and given the choice of entering a chamber smelling of OCT or an identical one smelling of BNZ. For control and consistency purposes, the experiment is done simultaneously in an identical apparatus B with the shock-associated and control odors reversed. We define a conditioned or "trained" fly as one that has chosen to go into the chamber filled with the control odor after given the choice for 90 seconds.

It is observed that following training, a large percentage of wild type flies choose to avoid the smell that was present when they received the electric shocks. The percentage is calculated as a normalized *performance index* (PI) where

$$PI = \frac{(trained - untrained)}{total} x100\%$$

Typical PI values for wild type flies were between 75 and 90% giving us confidence that the flies have learned to associate the stimuli. The general procedure described above is a typical associative learning Pavlovian conditioning paradigm for behavioral experiments appropriate for a variety of animals and more details can be found in the literature [101, 108]. One improvement that was discovered (and increased wild type PI to about 90%) was that it was possible to collect the flies without subjecting them to $CO_2$ anesthesia after conditioning by simply taping the chambers sharply so that the flies entered a test tube through a funnel by their own inertia. The mechanical shock associated with such tapping was shown to have no effect on the flies' PI while the (light) anesthesia immediately following conditioning as well as the presence of a naturally leaky $CO_2$ tank in the darkroom was known to compromise PI scores.

## 3.5 Controls

During physics experimentation, background measurements play a significant role in determining the signal. Similarly, a dominant theme in biological behavioral research of the type described here is that of a set of measurements collectively called controls. When results such as, for instance, a decrement in learning and memory exhibited by transgenic animals are presented, they must always be quantified with respect to the equivalent in the control (non-transgenic or wild type animal).

For our behavioral analyses to have any significance we had to first determine that the flies expressing foreign TAU or overexpressing native TAU were not affected in ways unrelated to learning and memory. Thus we had to assess the experimental flies' task-relevant sensory behaviors and olfactory acuity to an attractive odor in addition to testing their ability to feel and avoid electric shock and to detect and avoid aversive odorants. Experience-dependent non-associative behavior was also tested by examining the effects of *pre-exposure* to odorant plus shock since pre-exposure to electric shock and one odorant tends to decrease the animals' ability to perform well in associative learning tasks that depend on electric shock and another odorant.

We also had to determine that our flies were viable and no neurodegeneration took place as the result of expression of the transgenes. Note that in all these control experiments, the main principle is to test the transgenic against the control, for instance, an absolute decrement in PI is not indicative of a TAU-mediated effect if it is mirrored identically in control animals.

Finally, one must be certain that the observed effects are TAU specific, i.e. that other proteins that do not bind the microtubular cytoskeleton and are (over)expressed with the same drivers, result in no memory effects. This was done and was discussed in Section 3.4.

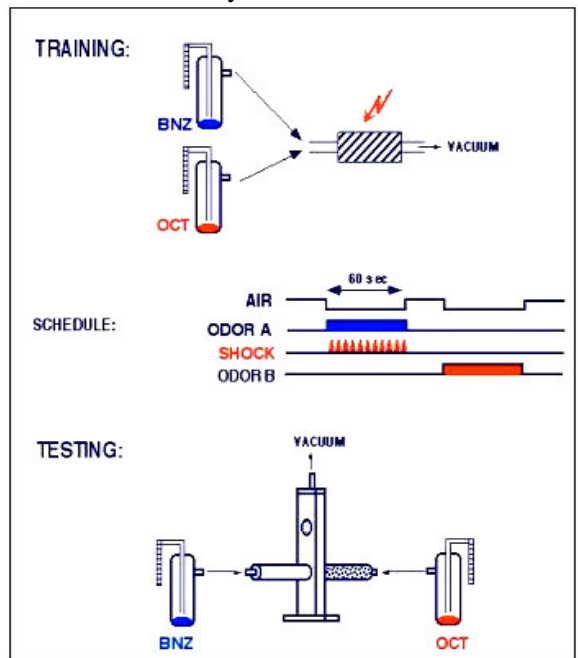

**Fig. 7** (Taken from [15]) Conditioning apparatus and training schedule.

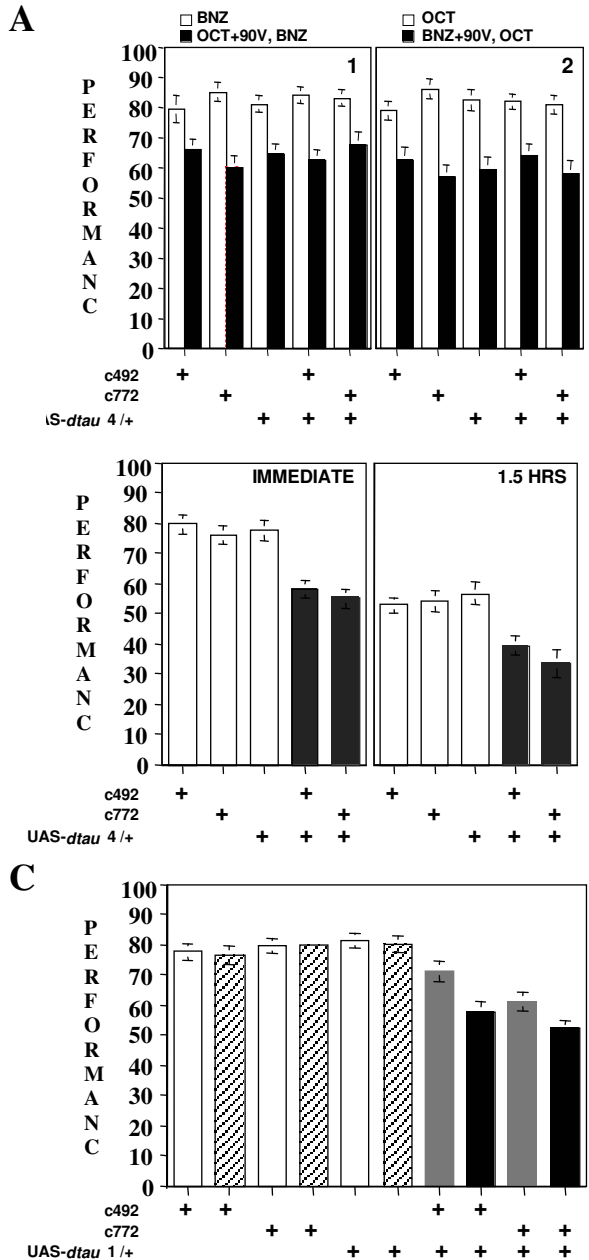

**Fig. 8 Results** (Taken from [9]) **(A)** Non-associative pre-exposure effect. 1. Avoidance of benzaldehyde after pre-exposure to full strength octanol and 90 V electric shock (filled bars) in comparison to avoidance without such pre-exposure (open bars) (n ≥ 7). ANOVA revealed significant effects of treatment ($F_{(1, 78)}$ =13.784, p<0.005, but not for genotype, both in pre-exposed and non pre-exposed animals. 2. Avoidance of octanol after pre-exposure to benzaldehyde and 90 V electric shock (filled bars) in comparison to octanol avoidance without pre-exposure (open bars) (n ≥ 7). ANOVA revealed significant effects of treatment ($F_{(1, 84)}$ = 14.026, p<0.005, but not for genotype, both in pre-exposed and non pre-exposed animals. **(B)** Olfactory memory after LONG paradigm conditioning. The mean Performance Index ± SEM of c492/+, c772/+, UAS-*dtau4*/+ (open bars) and c492/+; UAS-*dtau4*/+ and c772/+; UAS-*dtau4*/+ (filled bars) are shown (n ≥ 9). Two way ANOVA revealed significant effects of genotype [($F_{(4, 52)}$= 14.687, p<0.005) immediate (3-minute), and ($F_{(4, 49)}$= 9.327, p<0.005) 1.5 hours]. Subsequent Dunnett's tests for each time interval did not reveal significant differences in performance

among the c492/+, c772/+, UAS-*dtau4*/+ control strains or between the c492/+; UAS-*dtau4*/+ and c772; UAS-*dtau4*/+ heterozygotes. However, the differences between c492/+; UAS-*dtau4*/+ and c772; UAS-*dtau4*/+ heterozygotes and the control strains were highly significant (p<0.001) for immediate memory and 1.5 hr memories. **(C)** Performance of c492/+; UAS-*dtau1*/+ and c772; UAS-*dtau1*/+ heterozygotes with or without induction at 29°C after LONG conditioning. The average performance (PI$\pm$ SEM) of animals raised at 23-24°C for control strains (c492/+, c772/+, UAS-*dtau1*/+) is indicated by open bars and c492/+; UAS-*dtau1*/+ and c772; UAS-*dtau1*/+ heterozygotes with gray filled bars. The performance of animals raised at 23-24°C and subsequently induced at 29°C for 48-52 hours prior to behavioral experiments is indicated by the stippled bars for controls and the black-filled bars for c492/+; UAS-*dtau1*/+ and c772; UAS-*dtau1*/+ heterozygotes. Two way ANOVA indicated significant effects of genotype [($F_{(4, 44)}$= 8.287, p<0.005) for 23-24°C animals and ($F_{(4, 48)}$= 10.016, p<0.005) for animals induced at 29°C]. Subsequent Dunnett's tests revealed significant differences between the performance of c772; UAS-*dtau1*/+ heterozygotes and all control strains, as well as c492/+; UAS-*dtau1*/+ when uninduced (p<0.001). In contrast, both c772; UAS-*dtau1*/+ and c492/+; UAS-*dtau1*/+ heterozygotes were different than controls when the animals were induced at 29°C.

### 3.5.1 Mechanosensory

Sensory control experiments to ascertain that the transgenic flies retained their mechanosensory abilities (to feel and avoid pain caused by electric shock) were performed as described in [101, 108]. Avoidance of electrified grids kept at 92V (normal US stimulus), or 45V was not different between *tau*-expressing animals and controls, indicating lack of mechanosensory deficits due to TAU accumulation. The two levels of voltage were used as a further refining mechanism to help find any mechanosensory effects that our transgene expression may have had (e.g. it may have made the flies less sensitive but still able to sense 92V).

### 3.5.2 Olfactory Acuity
3.5.2.1 Pre-Exposure

Although, as expected CS+/US pre-exposure significantly reduced subsequent avoidance of the complementary odor, *tau*-expressing animals and controls exhibited equal decrements (Fig. 8A). Therefore, TAU accumulation did not cause differential responses to odor-shock pre-exposure and we can conclude that TAU accumulation does not affect experience-dependent, non-associative tasks.

3.5.2.2 Attractive and Aversive Odors

Both control and transgenic *tau*-expressing animals avoided equally the aversive odors benzaldehyde and 3-octanol (CS) at two different odor concentrations given the choice of fresh air. These results indicate that TAU accumulation in mushroom body and other central brain neurons described above, did not result in deficits in sensory abilities necessary for olfactory conditioning. In addition, we tested the response of *btau* and *htau*-expressing animals relative to controls to the attractive odor geraniol (GER) in olfactory trap assays [101].Though this odor is not task relevant, it provided an independent measure of olfactory acuity towards a qualitatively different odor.

As shown in Table 1 of [9] the performance of *tau*-expressing animals was not significantly different from controls for these tasks. Collectively, the results of these olfactory control experiments support the conclusion that despite the accumulation of TAU in antennal lobe neurons (that are used as olfactory sensors by the fly) *btau* and *htau* expressing animals retained their normal olfactory responses to the odors tested [9].

### 3.5.3 Viability of Transgenics

To assess the effects of extra TAU in MBs on the learning and memory capabilities of flies convincingly, we had to establish that our flies would be reasonably healthy during their conditioning and testing. We hypothesized that accumulation of TAU within the mushroom bodies would not affect survival because these neurons are dispensable for viability [109]. However, the effect on fly longevity of TAU accumulation in additional neurons where c772 and c492 drivers were determined to be active was unknown. In addition, it has been recently shown that expression of human wild type and mutant TAU proteins in the entire *Drosophila* nervous system (pan-neural expression), or targeted expression in cholinergic neurons, results in neurodegeneration and premature death of adult flies [96].In order to determine whether TAU expression in adult mushroom bodies and the other brain neurons described above affects the flies' viability, we evaluated the survival of *btau*, *htauwt1* and *dtau*-expressing flies over a period of 21 days post-eclosion[*].

Because both male and female animals are used for our behavioral experiments, we used mixed sex populations to evaluate survival unlike previous studies [96, 110]. We concluded that expression of vertebrate or *Drosophila tau* in adult mushroom body and other brain neurons did not result in decreased survival [9]. No overt differences between control strains and transgenics were observed for a limited set of animals that were evaluated for viability for two additional weeks (data not shown). Furthermore, TAU accumulation did not appear to result in gross morphological differences, or decreased fecundity and vigor from control strains.

These results indicate that TAU accumulation within the MB and other neurons of the adult brain does not precipitate the neurodegeneration-dependent decrease in survival observed with pan-neural expression of human mutant TAU proteins throughout development [96].

### 3.5.4 Neuroanatomy and Histology

Although the mushroom bodies are not essential for viability, we expected that degeneration of these neurons would severely impair behavioral neuroplasticity [109]. To determine whether TAU accumulation in adult mushroom bodies causes their degeneration, we histologically investigated the brain neuroanatomy of animals that expressed the *tau* transgenes. Because in past studies the severity of neurodegeneration was observed to increase with age and accumulation of TAU [109], we focused on 21 day-old animals. Using a semi-quantitative western blot [9] we concluded it is unlikely that any degeneration in older flies is the result of progressively increasing amounts of TAU in mushroom body neurons. Note here that although as previously mentioned, densitometry and immunohistochemistry are by nature hard to normalize, they do give reasonably accurate relative results. Thus we were able to determine that no more TAU was present in older flies even though we could not tell exactly how much overall TAU there was. Staining with various mushroom body antigenic markers [111] did not reveal detectable morphological anomalies in 21 day-old transgenic *tau*-expressing flies compared to similarly aged, or 2-3 day-old controls.

### 3.6 Results

The results of this study can be summarized as follows. Vertebrate (bovine and human) as well as native TAU accumulation in mushroom bodies (implying TAU bound to MTs) of adult flies results in associative olfactory learning and memory deficits.

### 3.6.1 Vertebrate TAU-Expressing Flies

We had determined that transgenic animals under the c772 MB driver express higher levels of TAU so we used c772/+; *htauwt1*/+ heterozygotes in the analysis presented below, and similar results were obtained for c4922/+; *htauwt1*/+ heterozygotes in a limited set of experiments (data not shown).

To determine whether TAU accumulation in the mushroom bodies affected associative processes, we trained *btau, htau-* and *dtau* expressing animals and controls in the long version of a negatively reinforced, olfactory associative learning task as described above. c492/+; UAS-*btauI*/+, c772/+; UAS-*btauI*/+ and c772/+; UAS-*htauwt1*/+ heterozygotes exhibited a highly significant 25-30% impairment in learning compared to controls (Fig. 8B, immediate). Similar results were obtained for *driver X dtau* and they are described in the literature [9]. These results demonstrated that TAU accumulation in the mushroom bodies strongly compromised behavioral neuroplasticity underlying associative olfactory learning and memory.

---

[*] Eclosion refers to the adult fly emerging from the pupal case (cocoon).

To more closely examine the learning and memory deficits of *btau, htau-* and *dtau* expressing animals, we utilized the SHORT variant of associative olfactory training [112] performed by E.M.C. Skoulakis. Because the LONG paradigm utilizes a 60-second CS+ presentation concurrent with 11, 92V electric shocks, the flies' performance represents learning from multiple rounds of what is referred to as 'massed' CS/US pairing. On the other hand, in the SHORT paradigm, a 10-second CS+ presentation is coupled to a single 1.25 second, 92V shock, allowing assessment after a single CS/US pairing [112, 113]. Furthermore, performance in SHORT training improves upon multiple pairings with a 15-minute inter-trial interval [112, 113]. This allows for a very fine-tuned experimental manipulation to produce equivalent learning in control and experimental animals, a necessary condition to investigate memory stability and retrieval properties.

The results in Fig. 8C demonstrate that a single CS/US pairing in *btau* and *htau-* expressing animals yielded losses in learning scores of the order of nearly 50% relative to controls. As with controls, the performance of *tau*-expressing animals improved upon multiple CS/US pairings indicating that the basic neuroplasticity mechanisms were at least operating in the right direction, but three CS/US pairings were necessary for *tau*-expressing animals to perform at the level reached by controls after only two pairings (Fig. 8C). This suggests that TAU accumulation causes either an impairment in the learning resulting by each CS/US pairing, or a compromise of memory stability, retrieval, or a combination of the two.

To distinguish between these three possibilities, we trained c772; UAS-*btau* and c772; UAS-*htauwt1* heterozygotes to the same performance level as controls (3 pairings for *tau*-expressing animals and 2 for controls) and measured memory of the association after 30-minutes. The *tau*-expressing animals exhibited a significant decrease in 30-minute memory, despite performing equivalently to controls immediately after training. This indicates that memory retrieval and /or stability were compromised in TAU-expressing animals. This result implicates TAU within mushroom body neurons to mechanisms that are key to memory stability and/or retrieval.

Since our coimmunoprecipitation experiments showed that all TAU tested did bind to MTs we can conclude that that the behavioral deficits observed *are the effect of burdening MTs with excessive TAU*. This is in accord with what one would expect if the MTs were the first (or at least near the 'front lines') of intercellular information manipulation elements

### 3.6.2    Integrating Results from h-, b-, dTAU Expressing Animals

Combining the results summarized above and those in [9] we are led to the conclusion that the decrements in learning and memory observed in *btau* and *htau*-expressing animals were not caused by accumulation of a vertebrate protein, but rather by increasing the level of TAU in these neurons. Low levels of *dtau* transcription did not affect the performance of c492/+; UAS-*dtauI*/+ animals. However, elevation of *dtau* transcription precipitated learning deficits similar to those observed with vertebrate *tau* and *dtau4* transgenics [9]. These results strongly indicate that the associative learning and memory deficits in vertebrate *tau* and *dtau* expressing animals are very likely the direct result of elevated TAU accumulation within mushroom body neurons and not because of the conformational differences between vertebrate and *Drosophila* proteins.

Finally, another -unlikely- scenario that fits the data is that since the native and overexpressed dtau  genes were in different parts of the genome, some role was played by the directed expression on the conformation of the extra dTAU resulting in a perturbation that made dTAU behave like its b and h analogues. One way to rule this out would be to entirely knock out the native TAU gene replace it with *htau* or *btau* thus testing whether these will take up the role meant for dtau thus possibly refining the findings reported here and also determining whether the hypothesis behind the "fetal, 4R" TAU versus the "adult 3R" TAU in Alzheimer's disease is correct.

## 3.7    Conclusions

Collectively, the results of the behavioral analyses suggest that the level of TAU within mushroom body neurons is essential for both olfactory learning elicited by each CS/US and memory retrieval or stability. The areas of significant homology between the vertebrate and *Drosophila* TAU are confined to the tubulin binding sites and the vertebrate protein appears to bind microtubules in a way similar to the way the fly protein does. Taken together, the results strongly suggest that excess TAU binding to the neuronal microtubular cytoskeleton causes mushroom body neuron dysfunction exhibited as learning and memory deficits. This also indicates that although excessive TAU may not result in (immediate or medium-term) neurodegeneration, it is sufficient to cause significant decrements in associative learning and memory that may underlie the cognitive deficits observed early in human tauopathies such as Alzheimer's.

## 3.8    Discussion

The pre-tangle state of elevated tau has been the topic of limited study in the past (e.g. loss of TAU in axons and elevation in the somatodendritic compartment of neurons prior to tangle formation shown in humans and animal models [114, 115] but the possible effect of this state on neuroplasticity and possibly consciousness (if one is willing to make the leap of faith and extend these findings from fly to human), had not been previously explored. Similarly, splicing mutations that increase the level of 4R (fetal) TAU are the hallmark of many human tauopathies[116, 117]. It has been argued that accumulation of unbound TAU and subsequent NFT formation in human tauopathies may be the result of conformational changes [118]. However, the conformational differences between dTAU and its vertebrate homologs did not appear important in affecting learning and memory deficits in our study, at least at the level of resolution we could obtain. In contrast, the overall level of TAU i.e. the MAP:MT stoichiometry within mushroom body neurons appeared to be of primary importance.

TAU accumulation in mushroom body neurons caused robust associative learning and memory deficits. It is surprising that within the resolution limits of our techniques, the deficits appeared confined to associative learning and memory and not to other experience dependent olfactory processes. These results suggest that normal cytoskeletal-mediated processes, likely disrupted by excess TAU are necessary for neuroplasticity undelying associative functions. One possible overarching explanation of our findings involves the disruption of axonal transport of vesicles and is explored fully in [9]. An alternative, of interest to us as explorers of the QCI involves the roles for neuronal microtubules and their dynamic interaction with the proper ratio of MAPs in learning and memory discussed in Section 2 and [7, 8], or as neuronal computational elements proposed in [16]. These data are consistent with specific predictions of these models, including the GSM model which predicts that perturbations in the ratio of microtubule-binding proteins will precipitate learning and memory dysfunction and also with the general approach behind the dipole-dipole logic suggestion where extra TAU would alter the local dielectric constant $\kappa$ by virtue of its increased density.

In summary, these results strongly suggest that the stoichiometry of TAU and microtubules within neurons is essential for behavioral neuroplasticity. Increasing the level of TAU within neurons precipitates deficits possibly due to inhibition of microtubule-dependent intraneuronal traffic, microtubule stability or interactive capacity. The strong behavioral effects indicate that directed TAU-accumulation within neurons can be used as a tool to disrupt and study neuronal function in general.

**SECTION 4:   REFRACTOMETRY, SURFACE PLASMON RESONANCE  AND DIELECTRIC SPECTROSCOPY OF TUBULIN AND MICROTUBULES**

## 4.1 Theory of Dielectrics
### 4.1.1 Dielectric Properties of Polar Molecules

The dipole moment $p$ is defined as a vector associated with a separation of two identical point charges. Its magnitude is defined as the (positive) charge times the displacement vector between the positive and the negative charge and its direction is from the negative to the positive

$$p = q \cdot d \qquad (27)$$

where q is the charge (in Coulombs) and $d$ the displacement vector pointing from - to +. Units of 'Debyes' are customarily used where 1D = $3.338 \times 10^{-30}$ Coulomb $\cdot$ meters. For $N$ dipoles, in volume V, we define the total electric polarization vector $P$ as the total electric dipole moment per unit volume

$$P = Np/V \qquad (28)$$

The (time invariant) displacement electric field $\mathbf{D}$ in isotropic media is defined as

$$D = \varepsilon E = \varepsilon_o E + P \qquad (29)$$

where $\varepsilon_o$ and $\varepsilon$ are the permittivities of free space and sample respectively and $E$ is the external electric field and thus

$$P = (\kappa \text{-} 1)\, \varepsilon_o E \qquad (30)$$

Where we have defined $\kappa = \varepsilon / \varepsilon_o$ the optical frequency *dielectric constant* of the material which is related to the refractive index n via

$$\kappa = n^2 \qquad (31)$$

The dielectric permittivity $\varepsilon$ of a substance is a measure of its ability to "neutralize" part of a static electric field by responding to it with a displacement of some of its localized charge. This charge displacement is referred to as polarization and is not dependent on a material having excess charge. Even for a static electric field but most importantly when the incident field is time-varying the dielectric permittivity will also depend on time. Because the capacitance (ability to store charge for a given potential difference) C of a medium is directly proportional to its $\varepsilon$ (as in the elementary case of the parallel plate capacitor where $C = \varepsilon A/d$ with $A$ the area and d the separation of the plates in the limit $d^2 << A$), $\varepsilon$ can be measured by inserting the medium between the plates of a capacitor and noting the ratio of the capacitance with ($C$) and without ($C_o$) the medium so that $\varepsilon = C/C_o$. This general basic principle holds even for a fluctuating field but with certain modifications as will be illustrated later.

Molecular electric polarizability $\alpha$ is a scalar of proportionality that quantifies the polarization of a sample as a result of application of an electric field which in general can have four components: electronic $\alpha_e$ (sensitive even to high frequency fields), ionic or atomic $\alpha_i$ (medium frequency), orientational or dipolar $\alpha_d$ (low frequency) and interfacial $\alpha_{dc}$ (very low to DC frequencies). For simplicity, we will assume that $\alpha$ is an isotropic characteristic of a protein solution sample, which is justifiable at low concentrations. The total polarizability as a function of frequency $\alpha(\omega) = \alpha_e(\omega) + \alpha_i(\omega) + \alpha_d(\omega) + \alpha_{dc}(\omega)$ is a good parameter to use when describing a system such as a protein in solution since, unlike the total dipole moment it does not change as a result of solvation, changes in pH, or local electric field ($E_{loc}$) amplitude or direction. We define the total dipole moment as the sum of the permanent dipole moment added to polarizability-dependent dipole moment.

$$p = p_{perm} + \alpha E_{loc} \qquad (32)$$

It can be shown [119] that a molecule in a spherical cavity surrounded by a medium of volume polarization $P$ will experience a local electric field

$$E_{loc} = E + P/3\varepsilon_o \qquad (33)$$

The above is known as the Lorentz field approximation and is applicable in the case of simple dipolar rotor molecules. Combining equations (28), (30) and (33) one finds that the average electric dipole moment to be

$$\vec{p}_{av} = \frac{\alpha}{3\varepsilon_o}\left(\frac{\kappa+2}{\kappa-1}\right)\vec{P} \tag{34}$$

where $\kappa=\varepsilon$=relative dielectric constant, and one arrives at the Clausius-Mossotti relation for electric polarizability

$$\alpha = \left(\frac{\kappa-1}{\kappa+2}\right)\left(\frac{3\varepsilon_o V}{N}\right) \tag{35a}$$

and by extension,

$$\vec{P} = \frac{\vec{p}N}{V}$$
$$\vec{P} = (\kappa-1)\varepsilon_o\vec{E} \tag{35b}$$
$$\Rightarrow \vec{p} = \frac{V}{N}(\kappa-1)\varepsilon_o\vec{E}$$

Note that equations 35a and 35b can give $\alpha$ and $p$ in terms of macroscopic, measurable parameters such as the dielectric constant $\kappa$ and volume and number of molecules in the sample while equation (34) is not easily applicable to an experimental determination of $p$ as it contains the difficult-to-measure $P$. However, (35a,b) are not applicable to cases where the local field cannot be approximated by the simple field assumed by equation (33). These cases include water where if the measured permanent dipole moment (6.1x10$^{-30}$ C x m) is inserted into the Clausius-Mossotti equation one arrives at a negative value for $\kappa$.

### 4.1.2 Dielectric in a Non-Polar Solvent

Noting that anti-parallel orientations between the local field and the dipole moment of a molecule in a sample will have higher interaction energies U (where U = - **p . E** = -|**p**||**E**|cosθ) than parallel ones, we can find the contribution of the permanent electric dipole of a molecule to the volume electric dipole moment of a bulk sample as follows: consider an equilibrium situation where the thermal energy $k_B T$ >> U and we can expand the probability of each dipole's orientation (determined by Boltzmann statistics $\propto \exp^{-U/k_B T}$) keeping only the zeroth and first-order terms. This gives the average dipole moment as

$$\mathbf{p}_{ave} = |\mathbf{p}|^2\mathbf{E}'/(3k_B T) \tag{36}$$

and the volume polarizability ignoring high frequencies can be written as

$$\alpha = \cancel{\alpha_e} + \alpha_i + \alpha_d = \alpha_i + |\mathbf{p}|^2/(3k_B T) \tag{37}$$

so volume polarizability varies inversely with temperature as expected (at lower temperatures it is easier to re-orient dipoles as $k_B T$ is closer to **p . E**). In the case of a dilute liquid, for instance when a protein is present at low concentration inside a non-polar buffer, and assuming the properties of the solution are the sum of the properties of the components, we can write the effective molar polarizations of each component in terms of their mole fractions. For instance, by replacing V/N in equation (35a) by $V_m$ the molar volume of the material, we can write the molar polarizability $\alpha_m = 3\varepsilon_o \left( \dfrac{\kappa - 1}{\kappa + 2} \right) V_m$ and also the molar polarization $P_m = \dfrac{a_m}{3\varepsilon_o} = P_o + P_p$ where

$P_o$ and $P_p$ are the induced and permanent dipole moments respectively. $P_o = \alpha_o / 3\varepsilon_o$ and $P_p = |\mathbf{p}|^2/(9\varepsilon_o \, k_B T)$. In this dilute approximation, the molar polarization for an N-component solution would be given by $P_m = \sum_{i}^{N} X_i P_{mi}$ where $X_I$ is the mole fraction of the $i$-th component and $P_{mi}$ is its polarization. By performing an experiment in a step-wise fashion where all the ingredients are added one at a time (e.g. starting with the buffer base added to a non-polar solvent) one can determine all the $P_{mi}$'s. In the case of a binary solution with non-polar solvent (i.e. one with only induced polarization) we can approximate the molar polarization of the solvent $P_{m1}$ as that of pure solvent and using only the first term of (11) arrive at: $P_{m1} = \dfrac{(\kappa_1 - 1)M_1}{(\kappa_1 + 2)\rho_1}$ where $M_I$ is the mass and $\rho_I$ the density of the solvent. Thus to determine the solute molar polarization at each concentration we can use:

$$P_{m2} = P_{m1} + (P_m - P_{m1})\left( \frac{1}{X_2} \right) \tag{38}$$

where $P_m$ stands for he measured 'bulk' molar polarization of the binary solution. Note here that according to this simplistic formalism one would expect the result of (38) to be independent of concentration since it is supposed to be uniquely determined by the molecular structure. However, it is frequently observed that the polarization increases as the concentration decreases due to significant solvent-solute interactions. As a result, it is customary to report the molar polarization extrapolated to an infinite dilution i.e. $P_{m2}$ as $\lim(X_2 \to 0)$ [60][3]. We address further limitations of this approach below.

Hedestrand's procedure [120] for determining polarizabilities of solutions is based on the above approach plus assumptions that the dielectric constant and the density of the solution are linear in the solute mole fraction, i.e. in our case $\kappa = \kappa_1 + aX_2$ and $\rho = \rho_1 + bX_2$ where a and b are the derivatives of the dielectric constant and density with respect to mole fraction. Substituting these into (38) one obtains:

$$P_{m2} = \left( \frac{3M_1}{(\kappa_1 + 2)^2 \rho_1} \right) a - \left( \frac{(\kappa_1 - 1)M_1}{(\kappa_1 + 2)\rho_1^2} \right) b + \left( \frac{(\kappa_1 - 1)M_2}{(\kappa_1 + 2)\rho_1} \right) \tag{39}$$

where $M_1$ and $M_2$ are the molar masses of the solvent and solute and $\rho_1$ is the density of the solvent. It can be seen then that we only need to know the limiting slopes of the $\kappa$ and $\rho$ versus mole fraction slopes to determine $P_{m2}$ and these we can obtain by measuring a number of solutions of increasing dilution.

### 4.1.3 Dielectric in Polar Solvent and Generalized Case

When dealing with a protein it is important to realize that modeling it as a dipolar rigid rotator is only justified in very specific cases. In general, proteins have many rigid dipoles, polar substituents such as backbone amides, polar side chains and C-termini. Although constrained to be part of the protein, these have significant freedom and can rotate and translate at low incident field frequencies to give very large dielectric constant to proteins. The generalized Kirkwood-Fröhlich theory gives a way to combine the high-frequency dielectric constant with the complicated dipolar contributions to obtain our desired static or DC dielectric constant of the protein in a polar solution. In this approach the sample of dielectric constant ε is approximated to a collection of permanent rigid dipoles embedded in surroundings of dielectric constant $\varepsilon_\infty$ which represents the sample's high frequency dielectric constant (which can be easily determined from a measurement of the refractive index). Focusing on a spherical region of volume V, of the order of the size of a molecule of sample and using classical continuum theory the effective aligning field is calculated ($E_{eff}$) as a function of the average field in the medium E which results in the following relation [121, 122]:

$$\frac{3V\varepsilon_o(\varepsilon - \varepsilon_\infty)(2\varepsilon + \varepsilon_\infty)}{\varepsilon(\varepsilon_\infty + 2)^2} = \left(\frac{\partial}{\partial E_{eff}}\left\langle \vec{M} \bullet \hat{E}_{eff} \right\rangle\right)_0 \tag{40}$$

where $\vec{M} = \sum_i \vec{p}_i$ is the total instantaneous dipole moment of the spherical volume V, the vector sum of the individual dipoles $\vec{p}_i$, and it is dotted into the unit vector pointing in the direction of $E_{eff}$. The angled brackets denote statistical thermodynamic average and the derivative is evaluated at zero field strength. Note that the $\vec{p}_i$ have the magnitude of the dipole moments in a theoretical 'gas' phase. A relationship between the thermal fluctuation term can be derived from statistical mechanics and one finally arrives [123] at:

$$\frac{3V\varepsilon_o(\varepsilon - \varepsilon_\infty)(2\varepsilon + \varepsilon_\infty)}{\varepsilon(\varepsilon_\infty + 2)^2} = \frac{\left\langle (\vec{M} \cdot \hat{E}_{eff})^2 \right\rangle_0 - \left\langle \vec{M} \cdot \hat{E}_{eff} \right\rangle_0^2}{k_B T} \tag{41}$$

by including the correction term g [121] in the right-hand side of the Kirkwood-Fröhlich theory $Ng\frac{p^2}{3k_BT}$ (N is the number density of dipole molecules, **p** as before the 'gas' phase moment of one molecule) one can account for the correlation between dipoles. If g = 1 moments are entirely free from interaction with each other. For g>(<)1 we have positive (negative) correlation, both physical in the case of polar liquids. So g is in a sense the ratio of actual fluctuations to theoretical 'gas' uncorrelated fluctuations and it has been shown [123] to be $g = \sum_{j=1}^{n_d} \left\langle \cos\theta_{ij} \right\rangle_0$ where $\theta_{ij}$ is the angle between dipoles **p**$_i$ and **p**$_j$ and $n_d$ is the number of dipoles in the sample. Gilson and Honig [123] generalized this to the case of proteins in polar or non-polar environments and *constrained* dipoles arriving at:

$$g' = \frac{\left\langle (\vec{M} \cdot \hat{E}_{eff})^2 \right\rangle_0 - \left\langle \vec{M} \cdot \hat{E}_{eff} \right\rangle_0^2}{\sum_{i=1}^{n_d}\left[\left\langle (\vec{p}_i \cdot \hat{E}_{eff})^2 \right\rangle_0 - \left\langle \vec{p}_i \cdot \hat{E}_{eff} \right\rangle_0^2\right]} \tag{42}$$

and also introduced a constraint factor C which accounts for the reduction of the freedom of individual dipoles

$$C = \frac{n_d p^2}{3\sum_{i=1}^{n_d}\left[\left\langle (\vec{p}_i \cdot \hat{E}_{eff})^2 \right\rangle_0 - \left\langle \vec{p}_i \cdot \hat{E}_{eff} \right\rangle_0^2\right]} \tag{43}$$

---

[3] This may be of relevance to some models of homeopathic drug action and help explain the unexpected solute aggregation in water upon dilution observed in [60].

So finally we have that:

$$\frac{3\varepsilon_o(\varepsilon - \varepsilon_\infty)(2\varepsilon + \varepsilon_\infty)}{\varepsilon(\varepsilon_\infty + 2)^2} = \frac{g'}{C} N \frac{p^2}{3k_B T} \tag{44}$$

As seen above, by measuring the dielectric constant of a tubulin solution at various concentrations (and incident frequencies) one can experimentally deduce the dipole moment of the free tubulin dimer. This is not straightforward since even though an ideal dielectric contains no free charge, parts of its constituent units (individual molecules, polymer filaments etc.) can suffer a localized separation of charge as a result of application of an external electric field

By using a solution with above-critical concentration it is also possible to monitor the changes in the dipole moment as tubulin dimers polymerize into MTs. At high frequencies where only $\alpha_e$ will contribute, one can use the Clausius-Mossotti equation with the substitution $\kappa = n^2$ and by measuring the refractive index of the solution arrive at the value of $|\mathbf{p}_{ave}|$.

## 4.2 Optics

The two basic laws of optics are the law of reflection: $\theta_i = \theta_r$ (the angle of incidence is equal to the angle of reflection) and the law of refraction, also known as Snell's law: $n_1\sin\theta_1 = n_2\sin\theta_2$ where $n_i$ refers to the refractive index of medium i and $\theta_i$ is the angle between the normal and the incident and refracted beams see Fig 9. Both these laws refer to specular processes (i.e. the boundary between the media can be assumed smooth) and can be derived from Maxwell's equations for electromagnetic waves incident on a boundary [119].

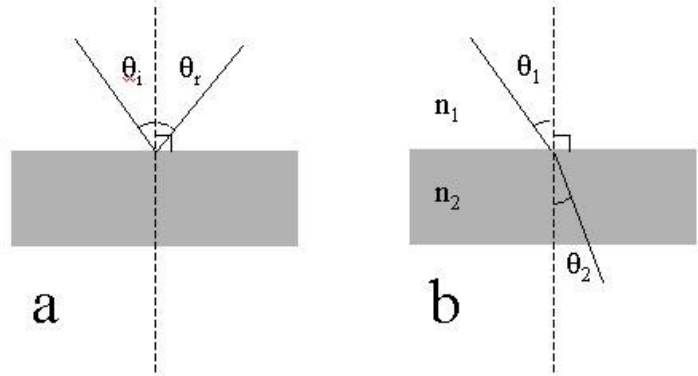

**Fig. 9 Laws of Optics. a)** Law of reflection **b)** Snell's law

Using Snell's law for $n_2 > n_1$ in Fig. 9(b) above, it can be easily shown that there will be a critical angle $\theta_c = \sin^{-1}(n_1/n_2)$ for a beam incident from $n_2$ such that the emergent beam will make an angle of $\theta_1 = 90^o$ and will be just grazing the surface. By exceeding this critical angle the beam is reflected back into the material (total internal reflection) and this is of importance in many applications including light guidance by optical fibers, sensing by surface plasmon resonance and refractometry. Note also that the refractive index is a slow function of temperature: $n_{T=25^{o}C} = n_T - (25.0 - T)(0.00045)$

## 4.2.1    Refractometry

The easiest way to measure n would be to pass a laser beam at some angle (other than the normal) through a sample of known thickness and measure the beam's deviation from the original path. This presents obvious problems with liquid sample containment etc. so practical refractometers are instead based on the phenomenon of total internal reflection and utilize high-quality prisms that can be tilted to compensate for wavelength-depended dispersion and thus ordinary sunlight or white light can be used as the source. The machine used for these experiments [Abbe Refractometer, Vista C10] was of exceptional accuracy and particularly elegant design requiring no electrical power. Briefly, light was allowed to enter and be reflected into a prism which was coated with the material of interest and covered. The prism's refractive index was known and the beam's incident angle was tilted until total internal reflection was reached (seen as a dark band in the eyepiece). The refractive index of the sample was then read on a pre-calibrated scale. This method depends on the prism having a higher refractive index than the material. Our machine was capable of measuring refractive indices between 1.3000 and 1.7000. After standard calibration and prior to measurement of tubulin solutions a number of different NaCl solutions of varying concentration were used as additional calibration. Determining the exact concentration was the main source of error in this measurement so high precision electronic scales and precision micro pipettes were used. The prism was cleaned after each measurement with ethanol soaked cotton and left to dry before applying the next sample. It was found that 30 to 50 µL of solution were adequate to deposit a thin film on the prism such that there was virtually no noise (indicated as colors). This small volume is comparable to the requirements of the sophisticated BIAcore 3000 SPR machine for a single injection.

The refractive index of a series of concentrations of NaCl (in 18.1MΩ/cm $H_2O$) and tubulin in buffer (0.1M 4-Morpholinoethane sulphonic acid (MES), 1mM EGTA, 0.1mM EDTA, 0.5mMMgCl$_2$, 1mM GTP at pH 6.4) was measured. Three measurements were taken for each data point and the average is shown in Table 1. Errors are estimated at ~5% for concentration (shown). The refractometer was scale limited with an error of 0.00005 for n (not shown, represented as the size of the data points. In Figures 10 and 11 a least-squares fit linear regression yields straight lines with R factors 0.9981 and 0.9928 respectively. The intercepts were manually set to the zero-point concentration averages. A limited second set of data points was taken ($n_{tub}2$) at several times after the first but as it was practically impossible to keep the timing consistent it is only shown here for qualitative purposes. The results at different times did not deviate appreciably suggesting that at this wavelength range tubulin dimers and microtubules have similar refractive indices.

| NaCl (mg/mL) | $n_{NaCl}$ | Tub (mg/mL) | $n_{tub}1$ | $n_{tub}2$ | time (min) |
|---|---|---|---|---|---|
| 0.000 | 1.3324 | 0.00 | 1.3352 | 1.3352 | 40 |
| 0.084 | 1.3321 | 0.00 | 1.3351 | 1.3353 | 51 |
| 0.167 | 1.3324 | 3.55 | 1.3415 | | |
| 0.208 | 1.3324 | 2.37 | 1.3386 | 1.3380 | 15 |
| 0.333 | 1.3325 | 1.18 | 1.3370 | 1.3371 | 17 |
| 0.416 | 1.3321 | 0.789 | 1.3364 | 1.3364 | 25 |
| 0.833 | 1.3323 | 0.592 | 1.3357 | 1.3360 | 25 |
| 1.67 | 1.3325 | 7.10 | 1.3478 | 1.3489 | 29 |
| 2.50 | 1.3329 | 7.10 | 1.3482 | | 7 |
| 3.33 | 1.3327 | 0.592 | 1.3357 | 1.3360 | 42 |
| 5.00 | 1.3335 | 1.183 | 1.3372 | 1.3370 | 45 |
| 6.67 | 1.3335 | 2.37 | 1.3390 | 1.3390 | 47 |
| 10.0 | 1.3341 | 0.473 | 1.3358 | 1.3358 | 55 |
| 20.0 | 1.3356 | | | | |
| 25.0 | 1.3364 | | | | |
| 33.3 | 1.3377 | | | | |
| 50.0 | 1.3406 | | | | |
| 100. | 1.3484 | | | | |

**Table 1 Refractometry Data.**

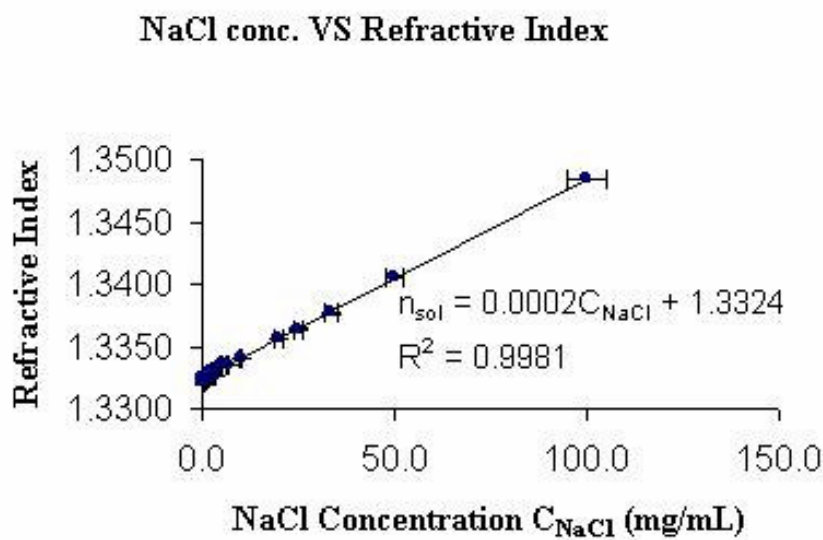

**Fig. 10 NaCl Concentration VS Refractive Index**. (Taken from [18])Note the low value of the slope (compared to tubulin).

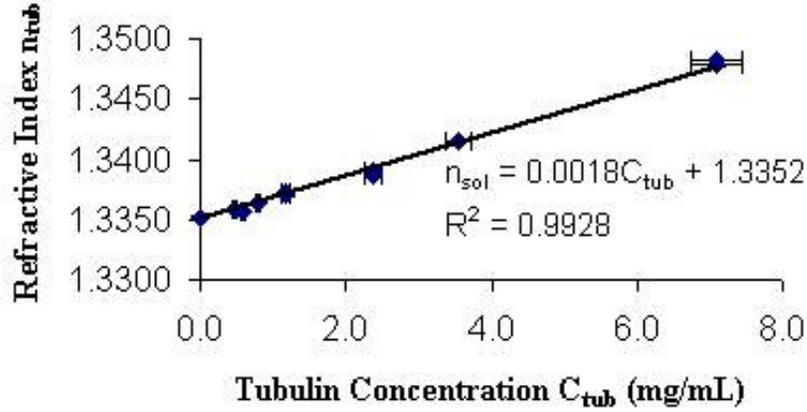

**Fig. 11 Tubulin Concentration VS Refractive Index.** (Taken from [18])

Note how a very small change in concentration of tubulin results in a large jump in the index of refraction (Fig. 11) giving a slope of $\Delta n/\Delta C = 1.800 \pm 0.090 \times 10^{-3}$ (strongly corroborated by the SPR measurement of the same value which gave $1.80 \pm 0.20 \times 10^{-3}$). Compare this to $\Delta n/\Delta C \sim 0.0002$ for the NaCl solution.

A physiological concentration is assumed at $15.0\mu M$ (i.e. $1.50 \times 10^{-5}$ mol/L) [124]. Since the molecular weight of a tubulin dimer is 110kD this gives a proportionality of $1.00$mg/mL $\sim 9.10$ $\mu M$ so $15.0\mu M$ is equivalent to $1.60$mg/mL and this gives a molecular density of $N = 9.03 \times 10^{18}$ tubulin molecules per liter or $9.03 \times 10^{21}$ per $m^3$. Note that the concentration necessarily varies across cell types, intracellular position and cell condition. For instance, although a TAU-burdened cell may have unchanged overall tubulin density, it has much higher local axonal and dendritic density of tubulin when MTs have deteriorated into neurofibrillary tangles.

The partial contribution to the refractive index of the solution by tubulin can be found if the density dependence on concentration of the solution is known (see Fig. 11). Assuming that the contributions from the various components are additive linearly, we have for the total index of refraction of the solution $n_{sol} = \sum_i C_i n_i$ where $C_i$ and $n_i$ are the fractional concentration of the $i_{th}$ component with refractive index $n_i$ and i runs over all components. Lumping the contribution from all the buffer components we can write

$$n_{sol} = (1 - \chi_{tub})n_{buffer} + \chi_{tub}n_{tub} \Rightarrow n_{tub} = \frac{n_{sol} - (1 - \chi_{tub})n_{buffer}}{\chi_{tub}} \qquad (45)$$

Where $\chi_{tub} = C_{tub}/\rho$ the mass fraction and $\rho$ is the density of the solution. It could be argued that using a volume fraction is more appropriate here but for our purposes a mass fraction is adequate and simpler. At $C_{tub} = 1.60$mg/mL, $n_{sol} = 1.8000 \times 10^{-3}C_{tub} + 1.3352$ (from Fig. 11) gives $n_{sol} = 1.34 \pm 0.07$ and using the measured value for the solution density $\rho = $ at $C_{tub} = 1.60$mg/mL we arrive at the value for $n_{tub}$

$$n_{tub} = 2.90 \pm 0.10 \qquad (46)$$

which can be used in

$$\kappa = n^2 \tag{47}$$

to give the high frequency tubulin dielectric constant

$$\kappa = 8.41 \pm 0.20 \tag{48}$$

A thorough search of the bibliography suggests that this is likely the first time these two quantities have been experimentally determined for tubulin.

Both $n$ and $\kappa$ are at the very top range of what is usually assumed for proteins, as expected since tubulin seems to have such a high dipole moment in molecular dynamics simulations.

The refractive index $n$ of an optically dense material is related to the high frequency polarizability $\alpha$ via:

$$n^2 = 1 + \frac{N\alpha}{\varepsilon_o} \frac{1}{\left(1 - \frac{N\alpha}{3\varepsilon_o}\right)} \tag{49}$$

where $N$ is the molecular concentration in which for our chosen concentration is $9.03 \times 10^{21}$ molecules/m$^3$. Solving the above equation for $\alpha$ gives

$$\alpha = \frac{\varepsilon_o}{N} \frac{3(n^2 - 1)}{(n^2 + 2)} \tag{50}$$

and therefore the high frequency tubulin polarizability is

$$\alpha = 2.1 \pm 0.1 \times 10^{-33} \qquad \text{C m}^2/\text{V} \tag{51}$$

A very large number owing to the evidently large dipole moment of tubulin. Note that the generally accepted value of the density of proteins is 1.45gr/mL [125].

## 4.3        Surface plasmon resonance (SPR)

### 4.3.1 SPR basics

The technique of surface plasmon resonance (SPR) [126, 127] allows measurement of changes in the optical properties of a medium adjacent to a thin metal film. Practical applications of the SPR method include chemical sensors [128, 129] and biosensors [130]. Specifically, the SPR technique is by now a well-established method for the analysis of interactions among biomolecules [131]. SPR curves (sensograms) can be measured either by varying the angle or the wavelength of the incident light [132-134]. Here, we discuss our application of the SPR technique to measurement of the dielectric properties of the bovine cytoskeletal protein tubulin.

A surface plasmon (SP) is an electromagnetic wave that can propagate along the surface of a dielectric-metal interface [127]. Surface plasmons can be excited by shining light on a layered system consisting of a transparent medium on one side, a metal film (most often gold or silver) and a dielectric on the other. When the light is incident at an angle greater than the critical angle of total internal reflection, an evanescent wave is produced and penetrates into the adjacent medium to a depth of the order of one wavelength. The maximum coupling between the evanescent wave and the surface plasmon takes place when their phase velocities coincide at which point the surface plasmon is excited at resonance. Thus, the surface plasmon resonance (SPR) occurs at a characteristic angle of incidence. This angle depends on the thickness as well as the dielectric permittivities of the layers of the adjacent media. Since the effective permittivities depend on the frequency of the exciting laser light, the resonance angle does too. The most convenient geometry for the development of a sensor is the Kretschmann-Raether configuration which consists of a glass prism, a metal film and the adjacent medium that is to be probed [135, 136].

### 4.3.2 The SPR Sensor

Fig. 12 shows the experimental arrangement of our custom-built SPR sensor [18, 135, 136]. In addition to the experimental arrangement shown, we used the commercial SPR-based

BIAcore 3000 and 1000 sensors, which furnished the additional convenience of automated injection of very small volumes of analyte solutions and allowed for cross-checking of results.

### 4.3.2.1 Optics and Data Acquisition

A helium-neon laser provided the incident illumination at 633 nm (or 760nm in the BIAcore). A p-polarized light beam convergent in an angular interval was produced with an arrangement of lenses and a polarizer. A prism provided the coupling of the laser beam to the SPs that were excited in the gold film, and a multiple-channel custom-build (see below) flow cell allowed solution access to the gold film. The angular distribution of the reflected light was measured with a photodiode array and its electrical output was read with a data acquisition (DAQ) board and transferred to a PC computer using software developed in-house by the author. The readout rate of the DAQ board set the time resolution to 67 milliseconds. The spatial resolution was determined by the dimensions of the laser beam spot at the surface: 0.5 mm×0.3 mm. The angular resolution for the configuration used was ~1 resonant unit (RU); this angular unit, commonly used in SPR measurements, corresponds to $10^{-4}$ degrees. The change of the SPR angle by this quantity occurred when the change of the refractive index was only approximately $10^{-6}$ [38].

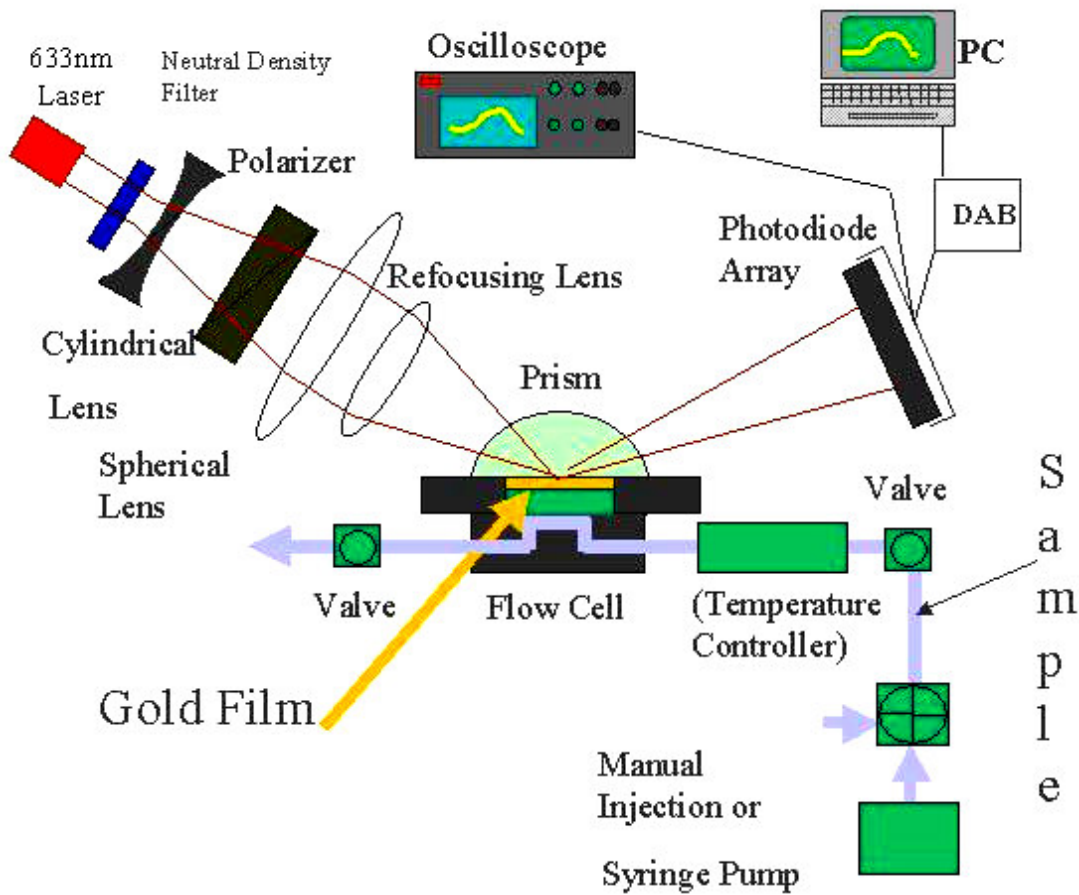

**Fig. 12 Experimental Setup of the SPR Sensor.** Sample flow is indicated with light blue. The prism provides the coupling of the excitation light to the surface plasmon. The polarizer is used to produce the p-polarized light, since only this component interacts with the surface plasmon. The angular distribution of the reflected light intensity is detected by the photodiode array which can be replaced by just two photodiodes with differential lock-in detection. The sample medium is injected into a small flow cell adjacent to the gold film. The fluidic cell presented a serious challenge and is described separately below.

### 4.3.2.2 Fluidic Cell

One of the serious challenges in an SPR measurement is the management and routing of the samples. In commercial devices, such as the BIAcore 3000, proprietary microfluidic technology based on lithography casting method is used and injections require robot arm handling and sophisticated electronics raising the cost of the device to several hundred thousand dollars. Previous methods used for in-house SPR measurements used hand-cut rubber and micro-machined Plexiglas cells and manual injection using microsyringes. In both rubber and Plexiglas cell designs, significant errors were introduced as a result of misalignment of the various cell parts, evaporation of sample and presence of air bubbles. In our custom-built apparatus, to bring a series of 30-100μL sample injections into close contact with a vertical gold film, avoiding air bubbles and mixing of samples, a casting technique was developed as follows. Using AutoCAD software, channels of precise dimensions (3mmx0.5mm and 3mmx0.3mm) were laser printed on thin plastic sheets (overhead transparencies). Those were then cut by hand under a dissecting stereoscope and pushed against modeling clay. The clay was pliable enough to follow the contours imposed by the plastic sheet positive and after oven curing became the negative mold.

Poly-Di-Methyl-Siliconate (PDMS) was used as the material for a fluidic cell. PDMS was found to be ideal for this application because it is biologically inert and virtually indestructible. Unfortunately, the material is non-machineable as holes drilled tend to self-seal but this was eventually advantageous for our purposes facilitating channel-syringe interfaces. None of the chemicals tried (ethanols, xylenes, acetone, hydrochloric acid) left any trace of abrasion, and freezing in liquid nitrogen and heating with an acetylene torch produced no effect. The material is transparent and we also found that this polymer is non-homogeneous in its fluorescence properties. Seemingly random parts of a large piece of PDMS would fluoresce strongly when exposed to UV laser illumination while others would be entirely transparent. This may be the result of domains of specially oriented polymer strands in the bulk of the material and would be interesting to investigate in a separate project. Stainless steel tubes (outer diameter 0.002, inner 0.001in) were used to interface with the channels and provided a way to inject and clear sample (see Fig. 13). The mold and PDMS were de-gassed in a low-vacuum (200mTorr) custom-built chamber and left to dry for 24 hours. The elasticity of the final product was of importance as if too elastic it would deform under pressure during sample injection and if too stiff if would not form the close contact needed to avoid evaporation. Elasticity was controlled by addition of curing and viscosity agents (Sylgard 186/184 and D.C. 200 respectively) and finding the optimum ratio.

The technique described above allowed us to comfortably have up to four independently-addressable channels while previously only one or at most two were possible. The limiting factor for the number of channels is the accuracy of the human hand in cutting the positives under a stereoscope. Delegating this task to a metal-deposition circuit board machine such as the T-Tech 700 would make it possible to obtain up to 8 channels before encountering problems with self-sealing.

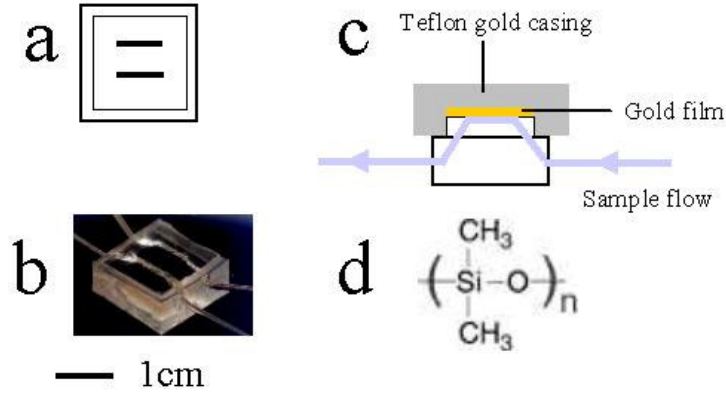

—— 1cm

**Fig. 13 PDMS Fluidic Cell. a)** Top view **b)** photograph of a two-channel cell **c)** side view **d)** PDMS chemical composition formula. The parenthesis is repeated n times (polymer)

### 4.3.2.3  Theoretical Model

To obtain quantitative data, we use a five-layer model that is based on Maxwell's equations describing reflection of light from a layered system. This enabled us to calculate SPR curves, estimate SPR response from protein immobilization and estimate changes in dielectric permittivity. We consider a structure with five layers: layer 1 consists of a prism with dielectric permittivity $\varepsilon_1$=2.30, layer 2 consists of a gold film of thickness $d_2$=47 nm with complex permittivity $\varepsilon_2 = -13.2 + i1.25$, layer 3 consists of a dextran layer filled with high relaxation (HR) solution of thickness $d_3$=140 nm [137] and $\varepsilon_3$=1.78 in the case of the sarcomeres, and with buffer in the case of tubulin, layer 4 consists of our sample medium (tubulin) with thickness $d_4$ and layer 5 consists of only solution. The fifth layer is assumed to be semi-infinite with respect to surface plasmon penetration depth. One can obtain the intensity reflection coefficient R for this system with the following recursive formula [138] by calculating each input impedance $Z_{in,m}$ and each layer impedance $Z_m$.

$$R = \left| \frac{Z_{in,2} - Z_1}{Z_{in,2} + Z_1} \right|^2 \tag{52}$$

The input impedance at layer m ($Z_{in,m}$) and layer impedance ($Z_m$) are obtained from,

$$Z_{in,m} = Z_m \left[ \frac{Z_{in,m+1} - iZ_m \tan(k_{z,m} d_m)}{Z_m - iZ_{in,m+1} \tan(k_{z,m} d_m)} \right], \quad \text{m=2,3,4} \tag{53}$$

where,

$$Z_{in,5} = Z_5, \quad Z_m = \frac{k_{Z,m}}{\varepsilon_m k_0}, \quad k_{Z,m} = \sqrt{\varepsilon_m k_0^2 - k^2}, \quad k = k_0 \sqrt{\varepsilon_1} \sin(\theta), \quad k_0 = \frac{2\pi}{\lambda} \tag{54}$$

The wavelength of laser in vacuum is $\lambda$. The incidence angle $\theta$ of light onto the prism-gold interface determines the component of the wave vector k that is parallel to the interface. A change in dielectric permittivity of sarcomeres or tubulin solution ($\varepsilon_4$) will alter the reflection coefficient R. Once the change in the SPR angle $\Delta\theta_{SPR}$ for different media is experimentally determined, the corresponding change in the dielectric constant ($\Delta\varepsilon_4$)can be calculated from equations (2) and (3) and the change in the refractive index n inferred from $\varepsilon = n^2$.

The decrease of the SPR sensitivity to changes in the dielectric permittivity with depth z starting from the boundary between $3^{rd}$ and $4^{th}$ layers into the protein sample was calculated. This decrease can be approximated as $\sim e^{-(z/d_p)}$ and the characteristic penetration depth $d_p$ can be calculated to be 110 nm [138]. For BIAcore, and $\lambda$=760nm $d_p \geq$ 110nm and therefore we do not expect to see saturation of the signal due to sensitivity degradation. This can also be seen if one considers that each free tubulin dimer occupies one binding site on the dextran surface thus saturation is achieved when the total mass of immobilized protein reaches that of a monomolecular layer (with thickness of $\leq$10nm).

### 4.3.2.4 Tubulin and Immobilization

Following established protocols [139], tubulin was purified from bovine cerebra (provided by R.F. Luduena). Our SPR measurements took place at 24$^o$C and the time between injection and measurement was of the order of 10 s. Measurements were taken for times up to 5 minutes. Tubulin does not polymerize at 0$^o$C and although 10sec is adequate time for our sample of 50 µL to reach room temperature and start polymerizing, we are confident that in our measurements mainly free tubulin dimers were present and not MTs since (using spectrophotometry and monitoring the absorption curve) we had previously determined that the characteristic time for our tubulin to polymerize into MTs was of the order of 45 minutes at room temperature (data not shown) agreeing with literature [140].

Gold film chips (CM5) coated with carboxymethylated dextran were obtained from BIAcore (BIAcore AB, Sweden). Using standard chemical activation/deactivation protocols [141], we introduced N-hydroxysuccinimide esters into the surface matrix of the chip by modifying the carboxymethyl groups with a mixture of N-hydroxysuccinimide (NHS) and N-ethyl-N'-(dimethyl-aminopropyl)-carbodiimide (EDC). These esters then form covalent bonds with the amine groups present in the ligand molecules thus immobilizing them on the surface. Effective immobilization requires that the pH be lower than the isoelectric point of the protein; however, lowering pH below 4.9 eliminates tubulin function and was avoided. The temperature controller raised and maintained the temperature of the inflow at 26$^o$C.

### 4.3.3 Surface Plasmon Resonance Results

In this setup, a 1ng/mm$^2$ surface immobilization yields a signal of 1kRU and the laser spot size on the gold chip is 1.2 mm$^2$. As shown in Fig. 14, after tubulin immobilization and application of a high flow rate (20 nL/min) of running buffer to wash away any weakly-bound protein, the average response was ~4 kRU which means 4.8 ng of protein (i.e. $2.6 \times 10^{10}$ individual tubulin dimers) were captured by the dextran. A reference cell on the same chip without any immobilized tubulin was used as control and any non-immobilization relevant signals (such as due to refractive index changes) were automatically subtracted.

As discussed earlier, electron crystallography measurements on zinc-induced tubulin protofilament sheets have shown that the tubulin heterodimer has dimensions 46 X 80 X 65 Å [17, 19, 20] so the footprint of the molecule on the surface can be between 30 nm$^2$ minimum and 52 nm$^2$ maximum depending on orientation. Using the average of these two values, it can be seen that a monomolecular layer covering the 1.2 mm$^2$ spot would require $3.0 \times 10^{10}$ individual tubulin molecules, leading us to believe that we have achieved 87% coverage, an observation corroborated by the immobilization part of the sensogram where a tendency towards saturation can be clearly seen (Fig. 14 D→E).

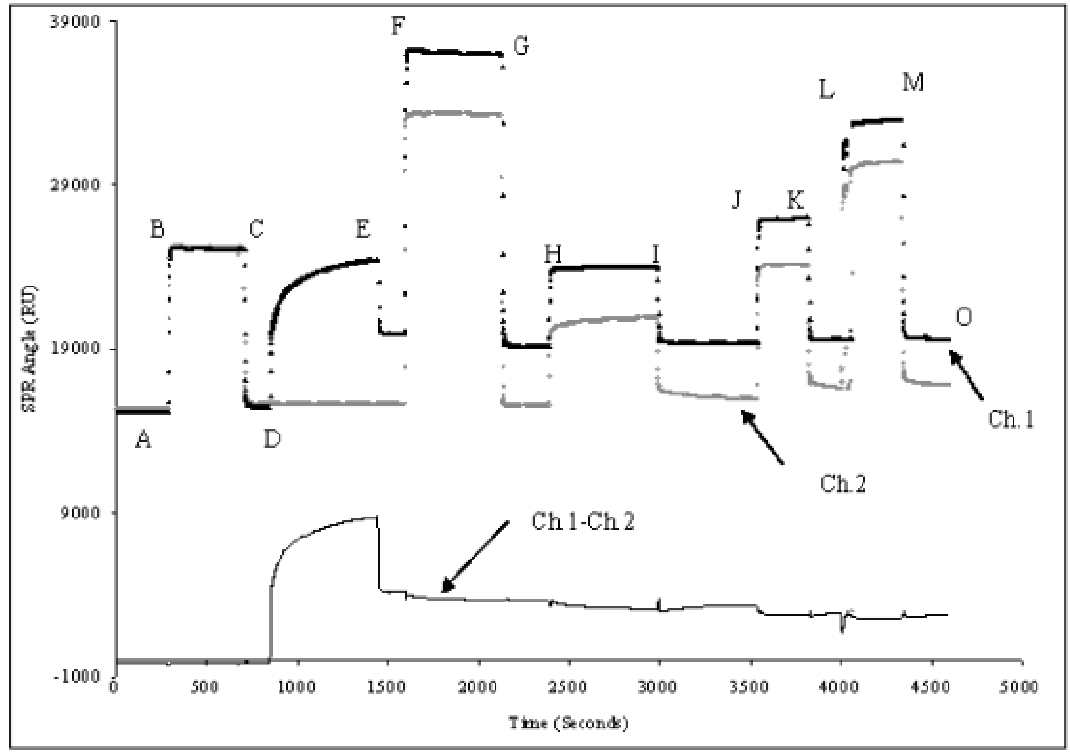

**Fig.14 Tubulin sensogram on BIAcore3000** (taken from [136]) Tubulin was immobilized on Channel 1. Channel 2 was treated identically to Channel 1 but had no tubulin. Ch.1 - Ch.2 shows tubulin signal - background. A→B: running buffer. B→C: EDC/NHS dextran-activating complex injected, note identical response of both channels. C→D: running buffer. D→E: Ch.1 shows tubulin immobilization with a clear tendency towards saturation, Ch.2 remains at running buffer baseline level. E→F: high flow rate running buffer. Difference between Ch1. and Ch.2 shows amount of immobilized tubulin ~4000RU. F→G: ethanolamine blocking (dextran-deactivation). G→H: running buffer. H→I: 0.51mg/ml tubulin in Ch.1. Since both channels exhibit same signal, all signal is due to refractive index change, not tubulin-tubulin binding (there is slight non-specific binding to Ch.2). I→J: running buffer. J→K: 1.70mg/ml tubulin in Ch.1, all signal is due to refractive index change. K→L: running buffer, L→M: 5.1mg/ml tubulin in Ch.1, all signal is due to refractive index change. Slight noise in the forms of bumps in Ch1-Ch.2 is due to 0.5 sec delay between measurement of Ch.1 and Ch. 2 and subsequent subtraction. Bump at around 4000sec is due to the temporary presence of a bubble in the 5.1mg/ml tubulin sample.

*In vitro* polymerization happens spontaneously at room temperature (also at 37°C only faster) if the protein concentration is above critical and the buffer contains adequate GTP. The critical concentration varies for different tubulin preparations. By using spectrophotometry, we determined that our tubulin started polymerizing at room temperature when concentration exceeded 1.0mg/ml (data not shown). In order to determine the dielectric constant of tubulin we first had to be sure that the shift in SPR angle was due to the change in the refractive index of the solution floating over the gold chip and not due to further immobilization of protein or perhaps tubulin-tubulin binding (polymerization).

To address the first concern, we performed the experiment in parallel, utilizing a reference channel on the same chip but without any tubulin in it. The reference signal was automatically subtracted from the tubulin signal thus also addressing concerns related to non-specific binding to deactivated dextran. To eliminate the possibility that our signal was due to further tubulin-tubulin interactions on the surface (polymerization) we tried both below critical (0.51mg/ml) and above critical (1.7mg/ml and 5.1mg/ml) concentrations and saw return to baseline in all cases showing that in this environment tubulin was incapable of polymerization, a fact that may be due to dextran binding and/or insufficient nucleation sites. Using the sensogram of Fig. 14 we calculated the changes of the refractive index and dielectric constant with tubulin concentration:

$$\frac{\Delta n}{\Delta c} = (1.85 \pm 0.20) \times 10^{-3} \left(mg \, / \, ml\right)^{-1} \Rightarrow \frac{\Delta \varepsilon}{\Delta c} = (5.0 \pm 0.5) \times 10^{-3} \left(mg \, / \, ml\right)^{-1} \tag{55}$$

where $\Delta n$ and $\Delta \varepsilon$ are the changes in the refractive index $n$ and dielectric constant $\varepsilon$; $\Delta c$ is the change in concentration $c$.

Since saturation occurs at only ~4kRU (nearly a monomolecular layer of tubulin), our assertion that we are dealing with free tubulin dimers is supported as each dimer must occupy one dextran binding site and there is no tubulin-tubulin binding or aggregation/polymerization into MTs.

As the dielectric constant and refractive index of a solution are intimately connected to the polarizability and consequently to the dipole moment of its constituents, these measurements show that SPR can be used to further elucidate the dielectric properties of 'live' proteins in solution.

### 4.3.4    Refractometry-SPR Comparison

Both refractometry and SPR, two methods based on the same underlying physical principles yet very far apart in implementation, gave a $\Delta n/\Delta c$ of $1.8 \times 10^{-3}$, a strong indication that these methods are consistent, our apparatus properly calibrated and our analysis correct. In summary, refractometry and SPR gave consistent results for the dielectric constant and polarizability of tubulin. These methods alone cannot provide the permanent dipole moment of the molecule since they address only the high frequency region where the permanent dipole is 'frozen out'.

## 4.4 Dielectric Spectroscopy

Earlier, we described obtaining the dielectric constant of tubulin at high frequencies. In order to probe lower frequencies we performed dielectric spectroscopy experiments.

### 4.4.1 Simplified Case

To illustrate the main principles behind the method used, consider the simplest way to measure the low-frequency dielectric constant of a solution: a capacitor-resistor (RC) circuit.

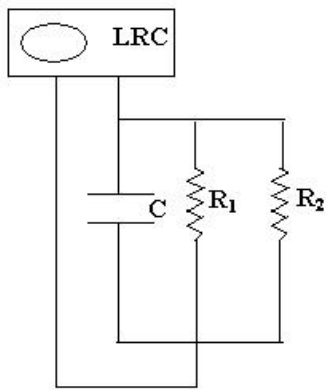

**Fig. 15 LRC Bridge.**

The capacitance C of a flow cell with conducting parallel-plate walls can be measured first filled with air and then filled with the tubulin-buffer solution for various tubulin concentrations. $R_1$ is assumed to be infinite in the case of air and some finite but very large value (of the order of 10M$\Omega$) in the case of solution. $R_2$ is set to a known value (e.g. 5k$\Omega$) and is needed to overwhelm any small conductance due to the presence of liquid between the plates. This ensures that the equivalent resistance in the circuit is nearly identical for all measurements. The inductance-resistance-capacitance (LRC) bridge performs the measurement at several low to medium frequencies (e.g. 1Hz to 32MHz) by measuring the RC time constant and displaying the total equivalent resistance and capacitance for the circuit. The ratio $C'/C$ for the tubulin/air capacitances then gives $\kappa$ and from equation (9) the dipole moment can be inferred. Unfortunately, a simple RC circuit with low inductance connected to a bridge such as the one described above and shown in Fig. 15 is not ideal for our case because aqueous solutions containing molecules with large dipole moments tend to form double layers at the electrodes giving extremely high values for $\kappa$ at low frequencies (the polarization effect). A further complication arises from the requirement that the sample volumes be as small as possible as purified proteins and especially tubulin are expensive.

### 4.4.2 Capacitor and Impedance Analyzer

We performed a dielectric spectroscopy measurement on liquid samples contained in a custom-made holder. A commercial impedance analyzer (Solartron 1260 from Solartron Inc.) was used that sampled the real and imaginary parts of the impedance from 1Hz to 32MHz. The device was calibrated with polar and non-polar molecules such as water, benzene and ethanol. By employing the distance variation polarization-removal technique it was possible to extract dielectric information in the lower limit of the frequency range even though individual scans were swamped by the electrode polarization effects below about 10kHz. The experimental set-up is shown in Fig. 16. Note that the sample holder had a guard ring to reduce fringe-field effects and allow use of a simple parallel-plate capacitor theoretical model.

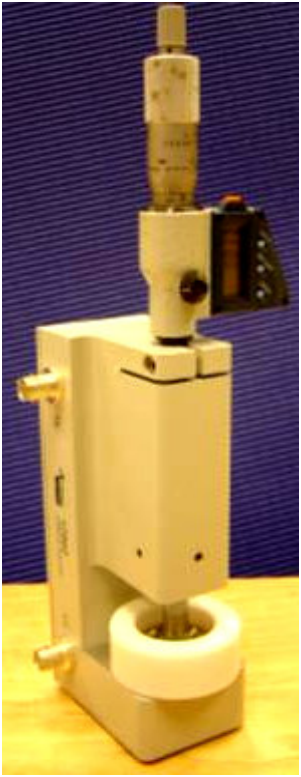

**Fig. 16** Capacitor cell. The capacitor had stainless steel plates 20mm diameter., with a precision digital micrometer controlling the plate separation (to 3μm precision), and a Teflon sample holder. Sample volume was ~1mL.

The impedance analyzer provides a sinusoidal voltage to the capacitor. As a result, alternating current $I$ exists in the sample and produces a voltage drop $V$ where $V=IZ$. Giving the impedance $Z$ as

$$Z = \frac{V}{I}.$$ (56)

The ratio of the output of channel 1(Hi) to that of channel 2 (Lo) is recorded as a function of frequency. The real and imaginary parts of the ratio were stored separately for additional processing.

The main source of error in measuring biological impedances is due to electrode polarization. This effect becomes very strong when measurements are carried out at low frequencies. We minimized this effect as follows. The measured impedance $Z^O$ is the sum of the sample impedance and the polarization impedance $Z^P$. We took two measurements of each sample at slightly different plate separations yielding $Z^O_1$ and $Z^O_2$ where,

$$Z^S_1 + Z^P = Z^O_1$$
$$Z^S_2 + Z^P = Z^O_2$$ (57)

Assuming the polarization effect to be the same in both cases (justifiable for similar separations), subtracting the two measurements gives

$$Z^S_1 - Z^S_2 = Z^O_1 - Z^O_2.$$ (58)

Our parallel plate capacitor had capacitance $C = \dfrac{\varepsilon A}{d}$, where ε is the dielectric permittivity of the suspension and A is the plate area and d is of the electrode separation and since the capacitor impedance is given by $Z = \dfrac{1}{Y} = \dfrac{1}{G + i\omega C}$, where G is the conductance and Y is the admittance so that $G = \dfrac{1}{R} = \dfrac{\sigma A}{d}$. From (58) we obtain

$$Z^O_1 - Z^O_2 = \frac{d_1 - d_2}{(\sigma + i\omega\varepsilon)A}$$ (59)

leading to the two dispersion curves:

$$\sigma(\omega) = \text{Re}(Z^O_1 - Z^O_2) = \text{Re}\frac{d_1 - d_2}{A(Z_1 - Z_2)}$$ (60)

$$\varepsilon(\omega) = \text{Im}(Z^O_1 - Z^O_2) = \text{Im}\frac{d_1 - d_2}{\omega A(Z_1 - Z_2)}$$ (61)

where σ(ω) and ε(ω) are the frequency dependent conductivity and permittivity of the sample respectively. It is thus seen that by obtaining the imaginary part of the transfer function, one can deduce ε(ω).

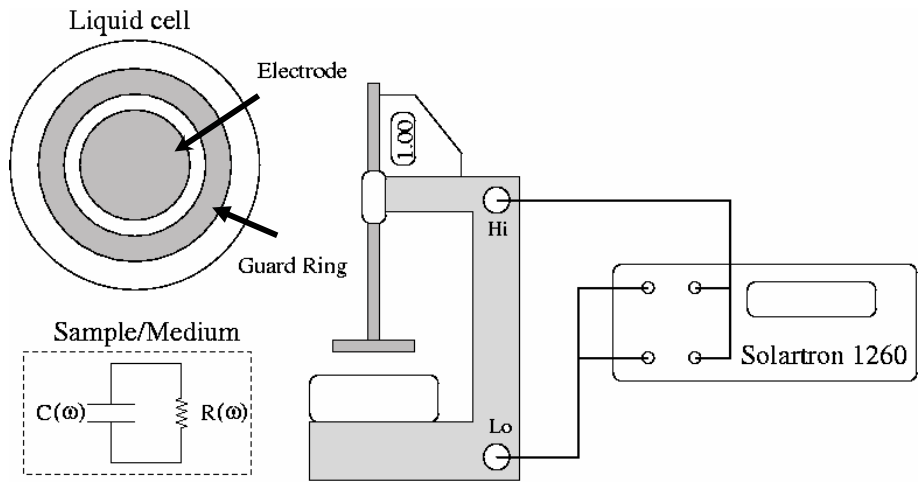

**Fig. 17 Experimental set up.** Solartron Dielectric interface was connected to the impedance analyzer. Sample holder had a guard ring to reduce fringe field effects.

### 5.4.3 Calibration and Errors

Several substances including polar and non-polar molecules were used for calibration as shown in Figs. 18 and 19.

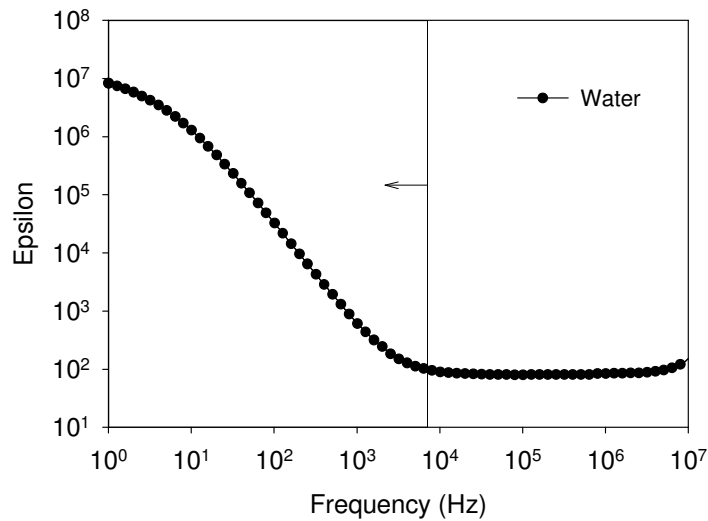

**Fig. 18 Water Calibration.** Pure, deionized water (18.0 MΩ/cm) tends to a relative permittivity of epsilon=κ ~ 80 at high frequency in accordance to values well established in literature. The sharp rise in the lower frequencies (shown by the arrow) illustrates the polarization effect.

Polar materials show a high polarization effect at frequencies higher than 10kHz, while gases and non polar liquids such as benzene do not show this error. We measured the dielectric constant of holder filled with plain air to corroborate that the obtained geometric capacitance has a κ of 1 as shown in Fig. 19.

Note that using a four-electrode technique would further reduce the polarization effect as would any electrode-less method if they could be adapted to small sample volumes.

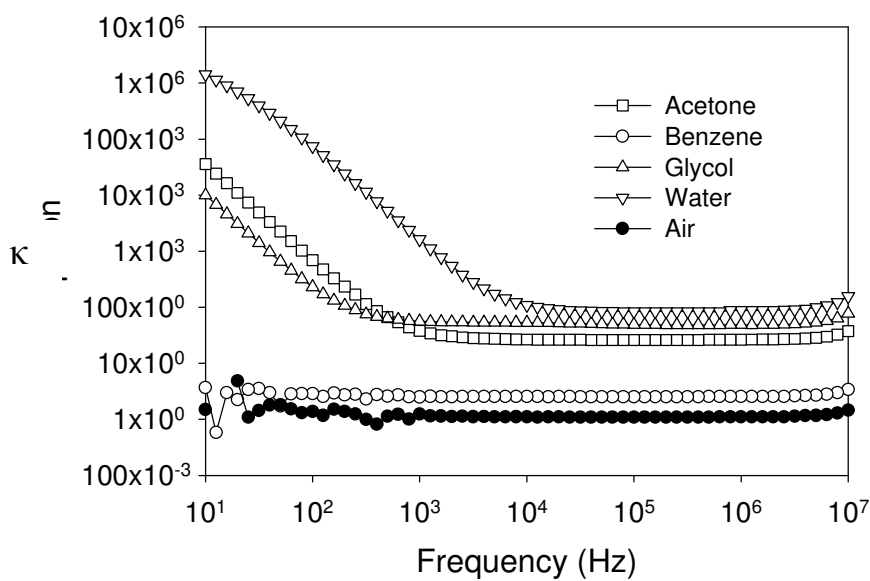

**Fig. 19 Dielectric spectrum of various substances.** For polar molecules like water and acetone we observed a strong polarization effect at frequencies lower than 10kHz, while for non-polar substances such as benzene with dielectric constant of κ=2.5 we do not observe this effect. Also for calibration we are able to read the dielectric constant of the air gap between the plates with κ=1.

### 4.4.4    Dielectric Spectroscopy Results

Purified tubulin of various concentrations was measured in standard buffer: 0.1M 2-Morpholinoethanesulfonic acid (MES) 1mM Ethylene glycol-bis(2-aminoethylether)-N,N,N',N'-tetraacetic acid (EGTA), 0.1 mM Ethylenediaminetetraacetic acid, 0.5mM MgCl$_2$ and 1mM of Guanosine 5'-triphosphate (GTP) . The results for concentrations of 0.1mg/ml and 1.1mg/ml. are summarized in Fig. 20 and 21 below.

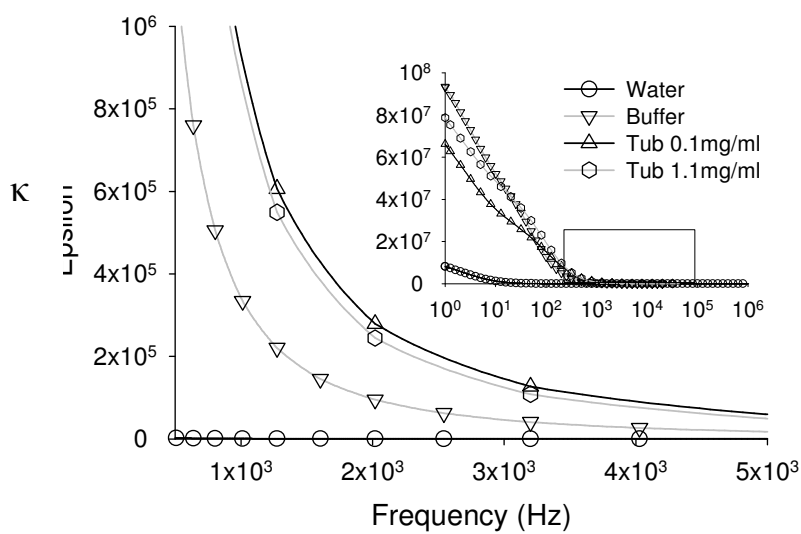

**Fig. 20 Combination Graph.** Composite graph of two indicative tubulin concentrations. The inset shows the entire frequency range behavior and the main graph is the close-up.

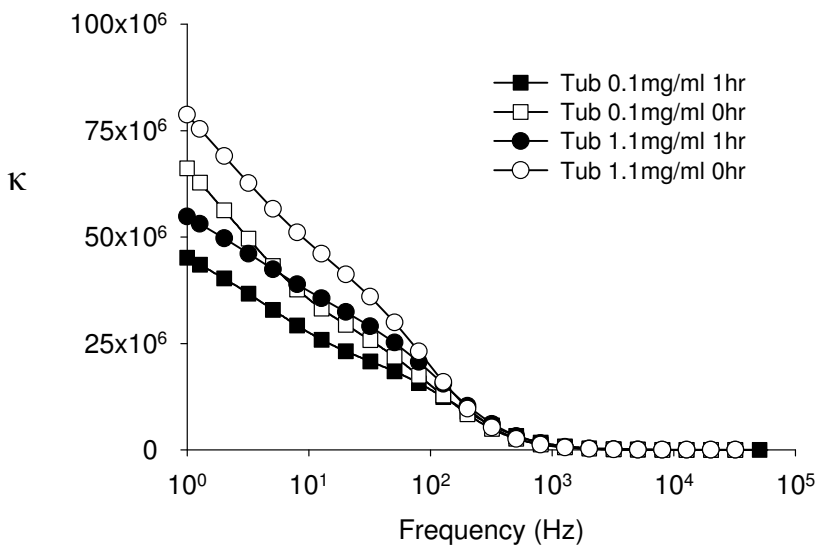

**Fig. 21 Time Dependence.** Time dependence of relative permittivity at different concentrations (□ 1.1mg/ml ○ 0.1mg/ml). Solid symbols correspond to t=1 hr while open symbols correspond to 0hr. after starting measurements. Differences may be attributable to polymerization of tubulin dimers into microtubules which is known to have a characteristic time of ~45min at room temperature.

Time dependence measurements were taken to show polymerization of tubulin dimers at concentrations higher than the critical concentration of 1mg/ml. We noticed some sedimentation of our sample at rates faster than the characteristic microtubule polymerization time.

### 4.4.5   Discussion of Dielectric Spectroscopy Results

It is stressed that currently, the preliminary dielectric spectroscopy results obtained by our group and summarized here are only shown as proof of principle. We have shown that it is possible, by taking data at a range of concentrations, to obtain the characteristic curves of concentration VS dielectric permittivity for tubulin and other samples of biological interest. In principle, by adding the components of a sample one by one it is possible to extract an accurate value for the dipole moment of a large protein molecule (and the rest of the components) at around 50kHz. Below this frequency a sophisticated error-reduction technique is required which is currently under development. From the preliminary data shown here, an order of magnitude calculation for the dipole moment of tubulin can be performed from equation (35) of this Section to yield |p| ~ $10^3$ Debye using the assuming no interaction of buffer and tubulin dipole moments. This value is purely qualitative but can be refined with further experimentation

## SECTION 5:   EMERGING DIRECTIONS OF EXPERIMENTAL TESTS OF THE QUANTUM CONSCIOUSNESS IDEA

### 5.1     Entanglement

Since 1935 when Erwin Schrödinger coined the word "entanglement" to refer to a state where the wavefunction describing a system is unfactorizable, much has been learned about this peculiar phenomenon and it has turned out to be very useful in quantum information science, quantum cryptography and quantum teleportation. Entanglement has been experimentally realized in light [142, 143], in matter [144] and in combinations of those [50, 145].One way to produce entangled states in light is via type II phase-matching parametric downconversion which is a process occurring when ultraviolet (UV) laser light is incident on a non-linear beta-barium borate (BBO) crystal at specific angles. A UV photon incident on a BBO crystal can sometimes spontaneously split into two correlated infrared (IR) photons (each of half the energy of the incident photon). The infrared photons are then emitted on opposite sides of the UV pump beam, along two cones, one of which is horizontally polarized and the other vertically. The photon pairs that are emitted along the intersections of the two cones have their polarization states entangled. This means that each photon is individually unpolarized, but the photons necessarily have perpendicular polarizations to each other. The state $\Psi$ of the outgoing entangled photons can be written as: $\left| \Psi \right\rangle = (\leftrightarrow, \updownarrow) + e^{i\alpha}(\updownarrow, \leftrightarrow)$ where the arrows indicate polarizations for the (first,second) IR photon and can be controlled by inserting appropriate half wave plates, while the phase factor $e^{i\alpha}$ can be controlled by tilting the crystal or using an additional BBO in a setup similar to the one depicted in Fig. 22, modified from [146].

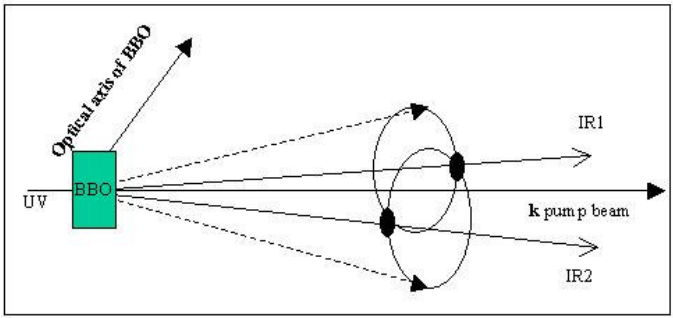

**Fig. 22 Type-II Phase-Matching Parametric Downconversion.** For certain orientations, a UV photon is absorbed by the BBO crystal and re-emitted as two entangled IR photons (IR1, IR2)

Measuring the state of one of the outgoing photons -say IR1, immediately determines the state of the other (IR2) regardless of their separation in space. This counterintuitive phenomenon is referred to as the Einstein Podolsky Rosen (EPR)-paradox and such pairs are called EPR pairs.

### 5.2        Molecular Electronics

Today's conventional silicon-based devices are of order 180nm in size while future molecular devices promise a further  order of magnitude reduction to this minimum. As a result, there have been considerable efforts concentrated on identifying various chemical substances with appropriate characteristics to act as binary switches and logic gates. For instance, rotaxanes have been considered as switches/fuses [38] and carbon nanotubes as active channels in field effect transistors [39]. Many of these substances are unsuitable for placement on traditional chips [40] or for forming networks, while virtually all of these efforts attempt to hybridize some kind of electrical wires to chemical substrates in order to obtain current flows. This complicates the task because of the need for appropriately nanomanufactured wires and connections.

We re-emphasize that the work presented here is suggestive of a different approach where the role of the binary states in an information encoding system is not played by the presence or bulk movement of charge carriers, but by naturally occurring conformational states of tubulin molecules and their self-assembled polymers -microtubules (MTs). Moreover, the external interaction with these states is to be performed by coupling laser light to specific spots of a microtubular network. Signal propagation is via traveling electric dipole moment flip waves such as those stipulated in Section 2 along MTs while modulation may be achieved by microtubule-associated-protein (MAP) binding that creates "nodes" in the MT network. In our proposed scheme for information manipulation, there is no bulk transfer of (charged) mass involved. Tubulin polymerization can be controlled by temperature and application of chemicals and MAPs to yield closely or widely spaced MTs, centers, sheets, rings and other structures [41, 42], thus facilitating fabrication of *nanowires, nodes and networks* and even structures capable of long-term information storage (*biomolecular computer memory*).

### 5.3 Proposed Further Research
### 5.3.1 On-Demand Entangled Photon Source

An on-demand entangled photon source (ODEPS) can be built and utilized to do spectroscopic analysis of proteins and other biomolecules. Such a device would greatly facilitate studies of the fundamental quantum properties of entangled objects, possible quantum properties of living matter, quantum information science and more. It would also provide an opportunity to test and identify the problems associated with the construction of a portable entanglement source to be taken "into the field" when entangled photons become indispensable to secure communications. For instance, tolerances to of entangled photon generation to vibration and temperature changes could be studied. A minimally modified design of the SPR apparatus developed for the work already done and presented here, coupled to a source of entangled photons would be capable of detecting the often conjectured mesoscopic bulk coherence and partial quantum entanglement of electric dipole moment states, existence of which will cast biomolecules as appropriate candidates for the implementation of *bioqubits*. One could follow a protocol similar to one developed by Oberparleiter et al. [147], capable of producing brightness in excess of 360,000 entangled photon pairs per second, coupled to a setup similar to the one developed by Altewischer et. al. [148] where entangled photons are transduced into (entangled) surface plasmons and re-radiated back as entangled photons. The essential difference would be that the insides of the perforations in the gold film of Altewischer et. al. would be covered with dextran to which a monolayer of tubulin dimers or microtubules would be immobilized by amine coupling (as described in detail in Section 4). The evanescent wave of the (entangled) surface plasmon generated at resonance will interact with the electric dipole moment of the immobilized protein complexes and presumably transfer the entanglement to a dipole state in a manner similar to the transfer of the photon polarization entanglement to surface plasmons. At the end of the tunnel, the surface plasmons would be reradiated having undergone the interaction with the protein electric dipole moment. If partial entanglement with the partner photon (that underwent none of these transductions) is found, then this would suggest that the protein is capable of "storing" the entanglement in its electric dipole moment state and characteristic decoherence times could be derived.

There are obvious objection to suggestions of long decoherence times for quantum properties of large molecules at room temperature and we have discussed possible ways to avoid early decoherence in Section 2. Here we note that these objections usually come from the application of equilibrium principles to the quantum mechanical aspects of the constituent atoms. We hope to investigate deeper, as although -for instance- the tubulin molecule consists of some 17,000 atoms which are subject to considerable thermal noise, the electric dipole moment state depends crucially on only a few electrons that can be in two sets of orbitals. In addition, tubulin is not an equilibrium system, rather it is a dynamic dissipative system where energy is being pumped in and out constantly. In fact, our theoretical work has suggested that for a certain set of parameters (such as the value of the dipole moment, the pH etc.) tubulin could indeed sustain a quantum mechanically coherent state for times of the order of microseconds [30].

### 5.3.2 Ways to Detect Quantum Coherence in Microtubules

In Section 2 and [30] comprehensive model conjecture treating certain regions inside MTs as isolated high-Q(uality) QED cavities was put forth as well as a scenario according to which the presence of ordered water in the interior of MTs results in the appearance of electric dipole quantum coherent modes, which couple to the unpaired electrons of the MT dimers via Rabi vacuum field couplings. The situation is analogous to the physics of Rydberg atoms in electromagnetic cavities [48]. In quantum optics, such couplings may be considered as experimental proof of the quantized nature of the electromagnetic radiation. In our case, therefore, if experimentally detected, such couplings would indicate the existence of coherent quantum modes of electric dipole quanta in the ordered water environment of MT, as conjectured in [56, 57], and used here.

To experimentally verify such a situation, one must first try to detect the emergent ferroelectric properties of MTs, which are predicted by this model and are potentially observable. Measurement of the dipole moment of the tubulin dimers is also an important step (see Section 4). A suggestion along these lines has been put forward in [12]. In addition, one should verify the aforementioned vacuum field Rabi coupling (VFRS), $\lambda_{MT}$, between the MT dimers and the ordered water quantum coherent modes. The existence of this coupling could be tested experimentally by the same methods used to measure VFRS in atomic physics [49], i.e. by using the MTs themselves as cavity environments, and considering tunable probes to excite the coupled dimer-water system. Such probes could be pulses of (monochromatic) light coupling to MTs. This would be the analogue of an external field in the atomic experiments mentioned above. The field would then resonate, not at the bare frequencies of the coherent dipole quanta or dimers, but at the Rabi-split ones, leading to a double peak in the absorption spectra of the dimers [49]. By using MTs of different sizes one could thus check on the characteristic $\sqrt{N}$-enhancement of the (resonant) Rabi coupling as described by equation 10 of Section 2 for MT systems with N dimers.

### 5.3.3 Other Ways to Measure Dipole Moments

Although the most direct approach to determine this quantity would be to measure the acceleration of evaporated single molecules with dipole moment in the gradient of an electric field in vacuum, this may prove a difficult task. Even though it is possible to keep tubulin from polymerizing in solution (e.g. by lowering the temperature and concentration), evaporating individual tubulin molecules may be very difficult due tubulin's affinity towards polymerization and aggregation. In other words the protein is naturally 'sticky' and will be hard to corpusculize. Furthermore, tubulin's electric dipole moment in vacuum is not directly relevant to its physiological value (although such a measurement would facilitate cross-check of theoretical molecular dynamics simulation calculations). In addition to the techniques described in detail above, the dipole moment can also be experimentally determined in a manner similar to the one used in "optical tweezers". A thin gold chip is bathed in a solution of purified depolymerized tubulin. A small spot ($1mm^2$) is illuminated with a continuous-wave laser beam of known wavelength and power. The laser beam's diameter can be modulated creating a gradient in the intensity of the beam. The dielectric moment of tubulin will interact with this gradient and the molecule will feel a ponderomotive electric force towards higher beam intensity. This, over a period of time, will concentrate the tubulin molecules at the center of the laser spot. The concentration of tubulin along the chip can be monitored in real time using a second laser beam exciting a SPR. Thus, by measuring the redistribution of molecules in response to the interaction with the intensity gradient and accounting for Brownian motion, one can evaluate the force exerted on tubulin and therefore the electric dipole moment.

## SECTION 6:   UNIFICATION OF CONCEPTS AND CONCLUSIONS

### 6.1  Putting It All Together

In our contribution to "The Emerging Physics of Consciousness", we have summarized roughly six years of theoretical and experimental work in physics, biophysics, biochemistry and neurobiology of tubulin and microtubules. Under the orchestration of Andreas Mershin (currently at the Center for Biomedical Engineering of the Massachusetts Institute of Technology) who did most of this work as part of his PhD at Texas A&M University (TAMU) Physics Department and the expert guidance of Dimitri V. Nanopoulos, we have organized a wide collaboration between experimental physicists (H.A. Schuessler's group at TAMU), theoretical physicists (the group of N.E. Mavromatos at King's College London), neurobiologists (the group of E.M.C. Skoulakis at TAMU and currently at the A. Fleming Institute in Greece) biochemists (the group of R.F. Luduena at the University of Texas, San Antonio) and dielectric spectroscopists (the group of J.H. Miller Jr. at the University of Houston) to study the notion that cytoskeletal proteins could be used in biomolecular based electronic and/or quantum information processing devices.

To this end and to aid in further theoretical study of tubulin and microtubules, our work concentrated on investigating the electric properties of this molecule. Supercomputer based molecular dynamics simulations were performed to study geometrical, energetic and electric properties of tubulin *in silico* determining the electric dipole moment (this computational answer agreed closely with similar simulations performed by others) and for the first time the dipole moments of the alpha and beta tubulin monomers were individually determined. In order to provide an experimental value to check the simulation results, precision *in vitro* refractometry with two independent methods (direct refractometry and surface-plasmon resonance sensing) was performed yielding the optical frequency polarizability and refractive index of tubulin for the first time. Furthermore, a custom-built dielectric spectroscopy apparatus was developed to demonstrate a method of obtaining the low (~50kHz) dielectric constant and as proof of principle a preliminary run was performed. This paves the way for more precise measurements of tubulin and other proteins and it is shown that the low frequency dielectric response of proteins can be used as an additional handle in proteomics as it depends strongly on the dipole moment of the protein. The latest theoretical approaches were described yet currently there is not an adequate theory to analytically and accurately describe the dipole-dipole interactions of a protein molecule in a polar solvent so only order-of-magnitude values for the dipole moment of tubulin were obtained based on simplifying assumptions. In order to establish a connection between the microtubular cytoskeleton and information processing *in vivo,* the effects of overexpression of microtubule associated protein TAU in the associative olfactory memory neurons of *Drosophila* were determined. Future directions of research concentrating on establishing the possible quantum nature of tubulin were developed theoretically.

Putting all of the above together makes for a wide scope study that impinged upon many aspects of tubulin. This work aims to show the way for focusing future efforts in a number of directions that will, hopefully, lead to a deeper understanding of proteins and the role that the dielectric and possibly quantum properties of biomolecules play in their function.

### 6.2. Conclusions

It has become increasingly evident that fabrication of novel biomaterials through molecular self-assembly is going to play a significant role in material science [149] and possibly the information technology of the future [142]. Tubulin, microtubules and the dynamic cytoskeleton are fascinating self-assembling systems and we asked whether their structure and function contain the clues on how to fabricate biomolecular information processing devices. Our work with the neurobiology of transgenic *Drosophila* [9] strongly suggests that the cytoskeleton is near the 'front lines' of intracellular information manipulation and storage. We also established that straight-forward spectroscopic techniques such as refractometry, surface plasmon resonance sensing and dielectric spectroscopy, coupled with molecular dynamic simulations and (quantum) electrodynamic analytical theory are useful tools in the study of electrodynamic and quantum effects in cytoskeletal proteins.

Implicit in our driving question is the possibility that if tubulin and MTs can indeed be made into the basis for a classical or quantum computer, then perhaps nature has already done so and tubulin and MTs already play such a role in living neural and other cells. If quantum mechanics is found to be important in cellular function (through its involvement in proteins) it is natural to ask whether there are indeed quantum effects in consciousness –if one reasonably assumes that the phenomenon of consciousness depends on cellular processes.

# ACKNOWLEDGEMENTS

The experimental work described here was undertaken mainly at Texas A&M University, the University of Houston and the University of Texas at San Antonio. We are indebted to the help and encouragement of Dr. Mita Desai of the National Science Foundation (NSF) and Prof. Jack A. Tuszynski. We wish to thank D. Chana, A. Michette, A.K. Powell, I. Samaras and E. Unger for discussions. We appreciate the technical support of Veena Prasad, Rita L. Williamson and Dr. Lisa Perez. The work presented here has been supported in part by NSF (grant No. 021895) and the Texas Informatics Task Force (TITF). In addition, AM was partially supported by the A.S. Onassis Public Benefit Foundation (Greece) and an Interdisciplinary Research Initiative grant from Texas A&M University. JHM is supported by the Texas Center for Superconductivity at the University of Houston, the Robert A. Welch Foundation (E-1221) and the Institute for Space Systems Operations (ISSO). RFL is supported by the Welch Foundation (grant No. AQ-0726). NEM is partly supported by the Leverhulme Trust (UK).